\documentclass[11pt]{book}
\usepackage{amsmath,epsfig,graphicx,amssymb}
\begin{document}
\setcounter{chapter}{1}

\begin{titlepage}

\centerline{{\Large {\bf Theory of superconductivity in doped
cuprates}}}

\vskip 15mm

\setcounter{page}{1}
\begin{center}
{\large {\bf Shiping Feng and Tianxing Ma}}\\
\vskip 5mm
{\large {\it Department of Physics, Beijing Normal University,
Beijing 100875, China}}\\
\vskip 15mm

\begin{minipage}{5in}
\centerline{{\sc Abstract}}
\medskip

Within the $t$-$t'$-$J$ model, the physical properties of doped
cuprates in the superconducting-state are discussed based on the
kinetic energy driven superconducting mechanism. We show that the
superconducting-state in cuprate superconductors is controlled by
both superconducting gap parameter and single particle coherence,
and then quantitatively reproduce some main features found in the
experiments on cuprate superconductors, including the doping
dependence of the superconducting gap parameter and
superconducting transition temperature, the electron spectral
function at $[\pi,0]$ point, the charge asymmetry of
superconductivity in the hole and electron doping, and the doping
and energy dependence of the incommensurate magnetic scattering at
both low and high energies and commensurate $[\pi,\pi ]$ resonance
at intermediate energy. We also show that the incommensurate
magnetic excitations at high energy have energies greater than the
superconducting gap energy, and are present at the superconducting
transition temperature.
\end{minipage}
\end{center}

{\bf Keywords:} Superconductivity; Single particle coherence;
Doped cuprates; Incommensurate magnetic scattering; Commensurate
resonance

\end{titlepage}
\newpage

\section{Introduction}

Since the discovery of high temperature superconductivity in doped
cuprates \cite{bednorz}, much effort has concentrated on the
superconducting mechanism \cite{anderson1,anderson2,laughlin}.
After intensive investigations over more than a decade, it has now
become clear that the strong electron correlation in doped
cuprates plays a crucial role not only for the unusual
normal-state behavior but also for the superconducting mechanism
\cite{anderson1,anderson2,laughlin}. The parent compound of
cuprates superconductors is a Mott insulator with the
antiferromagnetic long-range order, then changing the carrier
concentration by ionic substitution or increasing the oxygen
content turns these compounds into the superconducting-state
leaving the antiferromagnetic short-range correlation still intact
\cite{kastner}. Moreover, this superconducting-state is dependence
of both charge carrier gap function and single particle coherence
\cite{ding}. As a function of the charge carrier doping
concentration, the superconducting transition temperature
increases with increasing doping in the underdoped regime, and
reaches a maximum in the optimal doping, then decreases in the
overdoped regime \cite{tallon}.

In the conventional metals, superconductivity results when
electrons pair up into Cooper pairs, which is mediated by the
interaction of electrons with phonons \cite{bcs}. These electron
Cooper pairs condense into a coherent macroscopic quantum state
that is insensitive to impurities and imperfections in the
conventional metals and hence conducts electricity without
resistance \cite{bcs}. As a result, the pairing in the
conventional superconductors is always related with an increase in
kinetic energy which is overcompensated by the lowering of
potential energy \cite{bcs}. It has been realized that this
reduction in the electron potential energy actually corresponds to
a decrease in the ionic kinetic energy \cite{chester}, and thus
providing a clear link between the pairing mechanism and phonons.
In doped cuprates, the charge carriers form the Cooper pairs when
they become superconductors as in the conventional superconductors
\cite{tsuei}, which is supported by many experimental evidences,
including the factor of $2e$ occurring in the flux quantum and in
the Josephson effect, as well as the electrodynamic and
thermodynamic properties \cite{tsuei}. Although a possible doping
dependent pairing symmetry has been suggested \cite{yeh}, the
charge carrier Cooper pairs have a dominated d-wave symmetry
around the optimal doping \cite{tsuei,yeh}. By virtue of
systematic studies using the nuclear magnetic resonance, and muon
spin rotation techniques, particularly the inelastic neutron
scattering, it has been well established that the
antiferromagnetic short-range correlation coexists with the
superconducting-state in the whole superconducting regime
\cite{yamada,dai,wakimoto}. Furthermore, the unusual
incommensurate magnetic excitations at high energy have energies
greater than the superconducting pairing energy, are present at
the superconducting transition temperature, and have spectral
weight far exceeding that of the commensurate resonance at
intermediate energy \cite{hayden}. These experimental results
provide a clear link between the charge carrier pairing mechanism
and magnetic excitations, and also is an indication of the
unconventional superconducting mechanism that is responsible for
the high superconducting transition temperatures \cite{anderson1}.
In this case, it has been argued based on the non-Fermi liquid
normal-state that the form of the charge carrier Cooper pairs is
determined by the need to reduce the frustrated kinetic energy of
charge carriers \cite{anderson2}, i.e., the strong frustration of
the kinetic energy in the normal-state is partially relieved upon
entering the superconducting-state, where the driving attractive
force between holes may be attributed to the fact that by sharing
a common link two holes minimize the loss of the energy related to
breaking antiferromagnetic links, and is therefore mediated by the
exchange of spin excitations \cite{dagotto}. Moreover, by solving
a model for alkali doped fullerenes within dynamical mean-field
theory, it has been argued \cite{capone} recently that the strong
electron correlation does not suppress superconductivity, but
rather seems to favor it because the main ingredient was
identified into a pairing mechanism not involving the charge
density operator, but other internal degrees of freedom, like the
spin, unveiling a kind of the charge-spin separation. The
normal-state above the superconducting transition temperature
exhibits a number of anomalous properties which is due to the
charge-spin separation \cite{anderson1,anderson2}, while the
superconducting-state is characterized by the charge-spin
recombination. These scenarios are consistent with recent optical
experiments \cite{molegraaf}.

Recently, we \cite{feng1} have developed a charge-spin separation
fermion-spin theory to study the physical properties of doped
cuprates, where the electron operator is decoupled as a gauge
invariant dressed holon and spin. Within this theoretical
framework, we have discussed the unusual normal-state properties
of the underdoped cuprates and kinetic energy driven
superconducting mechanism, and the results are qualitatively
consistent with the experiments. It is shown that the charge
transport is mainly governed by the scattering from the dressed
holons due to the dressed spin fluctuation, while the scattering
from the dressed spins due to the dressed holon fluctuation
dominates the spin response \cite{feng1}. Based on the $t$-$J$
model, it is also shown \cite{feng2} that the dressed holons
interact occurring directly through the kinetic energy by
exchanging the dressed spin excitations, leading to a net
attractive force between the dressed holons. Then the electron
Cooper pairs originating from the dressed holon pairing state are
due to the charge-spin recombination, and their condensation
reveals the superconducting ground-state, where the maximal
superconducting transition temperature occurs around the optimal
doping, and then decreases in both underdoped and overdoped
regimes \cite{feng3}. However, the simple $t$-$J$ model can not be
regarded as a comprehensive model for the quantitative comparison
with cuprate superconductors. It has been shown \cite{well} from
the angle resolved photoemission spectroscopy experiments that
although the highest energy filled electron band is well described
by the $t$-$J$ model in the direction between the $[0,0]$ point
and the $[\pi,\pi]$ point (in units of inverse lattice constant)
in the momentum space, but both the experimental data near
$[\pi,0]$ point and overall dispersion may be properly accounted
by generalizing the $t$-$J$ model to include the second- and
third-nearest neighbors hopping terms $t'$ and $t''$. Moreover,
the additional second neighbor hopping $t'$ may play an important
role in explaining the difference between the electron and hole
doping \cite{feng4}. In this chapter, we discuss the physical
properties of cuprate superconductors within the framework of the
kinetic energy driven superconducting mechanism. We have performed
a systematic study within the $t$-$t'$-$J$ model based on the
charge-spin separation fermion-spin theory, and quantitatively
reproduced some main features found in the experiments on cuprate
superconductors, including the doping dependence of the
superconducting gap parameter and superconducting transition
temperature \cite{tallon}, the electron spectral function at
$[\pi,0]$ point \cite{well}, the charge asymmetry of
superconductivity in the hole and electron doping, and the doping
and energy dependence of the incommensurate magnetic scattering at
both low and high energies \cite{hayden,arai,tranquada} and
commensurate $[\pi,\pi ]$ resonance at intermediate energy
\cite{bourges0,bourges}. Our results also show that the effect of
the additional second neighbor hopping $t'$ is to enhance the
d-wave superconducting pairing correlation, and suppress the
s-wave superconducting pairing correlation.

This chapter is organized as follows. The theory of
superconductivity in doped cuprates is presented in Sec. 2. It is
shown that the superconducting-state in cuprate superconductors is
controlled by both superconducting gap function and single
particle coherence. In particular, this superconducting-state is
the conventional Bardeen-Cooper-Schrieffer like, so that some of
the basic Bardeen-Cooper-Schrieffer formalism \cite{bcs} is still
valid in discussions of the superconducting transition temperature
and electron spectral function, although the pairing mechanism is
driven by the kinetic energy by exchanging dressed spin
excitations, and other exotic properties are beyond
Bardeen-Cooper-Schrieffer theory. In Sec. 3, we calculate
explicitly the dynamical spin structure factor of cuprate
superconductors in terms of the collective mode in the dressed
holon particle-particle channel, and give a quantitative
explanation of inelastic neutron scattering experiments on cuprate
superconductors \cite{hayden,bourges0,bourges,arai}. In Sec. 4, we
discuss the charge asymmetry of superconductivity in hole and
electron doping, and show that the maximum achievable
superconducting transition temperature in the optimal doping in
the electron-doped case is much lower than that of the hole-doped
side due to the electron-hole asymmetry. Sec. 5 is devoted to a
summary.

\section{Theory of superconductivity}

Since the characteristic feature in doped cuprates is the presence
of the two-dimensional CuO$_{2}$ plane \cite{kastner}, then it
seems evident that the relatively high superconducting transition
temperature is closely related to doped CuO$_{2}$ planes. It has
been argued that the essential physics of the doped CuO$_{2}$
plane is well described by the $t$-$t'$-$J$ model on a square
lattice \cite{anderson1,well},
\begin{eqnarray}\label{E1}
H&=&-t\sum_{i\hat{\eta}\sigma}C^{\dagger}_{i\sigma}
C_{i+\hat{\eta}\sigma}+t'\sum_{i\hat{\tau}\sigma}
C^{\dagger}_{i\sigma}C_{i+\hat{\tau}\sigma}+\mu\sum_{i\sigma}
C^{\dagger}_{i\sigma}C_{i\sigma}\nonumber \\
&+&J\sum_{i\hat{\eta}}{\bf S}_{i} \cdot {\bf S}_{i+\hat{\eta}},
\end{eqnarray}
supplemented by the on-site local constraint,
\begin{eqnarray}\label{E2}
\sum_{\sigma}C^{\dagger}_{i\sigma} C_{i\sigma} \leq 1,
\end{eqnarray}
to avoid the double occupancy, where $\hat{\eta}=\pm\hat{x},\pm
\hat{y}$, $\hat{\tau}= \pm\hat{x}\pm\hat{y}$,
$C^{\dagger}_{i\sigma}$ ($C_{i\sigma}$) is the electron creation
(annihilation) operator, ${\bf S}_{i}=C^{\dagger}_{i}{\vec\sigma}
C_{i}/2$ is spin operator with ${\vec\sigma}=(\sigma_{x},
\sigma_{y},\sigma_{z})$ as Pauli matrices, and $\mu$ is the
chemical potential. The $t$-$J$ type model was originally
introduced as an effective Hamiltonian of the large-U Hubbard
model \cite{anderson1}, where the on-site Coulomb repulsion U is
very large as compared with the electron hopping energy $t$, which
leads to that electrons become strongly correlated to avoid double
occupancy, while the origin of $t'$ has been discussed
theoretically \cite{nazarenko}, and it appears naturally in
mappings from the three-band model for doped cuprates to a
one-band $t$-$J$ type Hamiltonian. Therefore the strong electron
correlation in the $t$-$t'$-$J$ model manifests itself by the
electron single occupancy local constraint, which is why the
crucial requirement is to impose this electron local constraint
for a proper understanding of the physical properties of doped
cuprates. To incorporate this local constraint, the charge-spin
separation fermion-spin theory \cite{feng1} has been proposed,
where the constrained electron operators are decoupled as,
\begin{eqnarray}\label{E3}
C_{i\uparrow}= h^{\dagger}_{i\uparrow} S^{-}_{i}, ~~~~~
C_{i\downarrow}=h^{\dagger}_{i\downarrow}S^{+}_{i},
\end{eqnarray}
where the spinful fermion operator $h_{i\sigma}=
e^{-i\Phi_{i\sigma}}h_{i}$ describes the charge degree of freedom
together with some effects of the spin configuration
rearrangements due to the presence of the doped hole itself
(dressed holon), while the spin operator $S_{i}$ describes the
spin degree of freedom (dressed spin), then the electron on-site
local constraint for the single occupancy,
\begin{eqnarray}\label{E4}
\sum_{\sigma}C^{\dagger}_{i\sigma}C_{i\sigma}&=& S^{+}_{i}
h_{i\uparrow} h^{\dagger}_{i\uparrow}S^{-}_{i}+ S^{-}_{i}
h_{i\downarrow} h^{\dagger}_{i\downarrow}S^{+}_{i} \nonumber \\
&=&h_{i} h^{\dagger}_{i}(S^{+}_{i}
S^{-}_{i}+S^{-}_{i}S^{+}_{i})=1- h^{\dagger}_{i}h_{i}\leq 1,
\end{eqnarray}
is satisfied in analytical calculations, and the double spinful
fermion occupancy,
\begin{eqnarray}\label{E5}
h^{\dagger}_{i\sigma} h^{\dagger}_{i-\sigma}=e^{i\Phi_{i\sigma}}
h^{\dagger}_{i} h^{\dagger}_{i}e^{i\Phi_{i-\sigma}}=0, ~~~~~
h_{i\sigma}h_{i-\sigma} =e^{-i\Phi_{i\sigma}}h_{i}h_{i}
e^{-i\Phi_{i-\sigma}}=0,
\end{eqnarray}
are ruled out automatically. It has been shown that these dressed
holon and spin are gauge invariant \cite{feng1}, and in this
sense, they are real \cite{laughlin}. At the half-filling, the
$t$-$t'$-$J$ model is reduced to the antiferromagnetic Heisenberg
model, where there is no charge degree of freedom, and then the
real spin excitation is described by the spin operator $S_{i}$.
Since the phase factor $\Phi_{i\sigma}$ is separated from the bare
spinon operator, and then it describes a spin cloud
\cite{feng1,martins}. Therefore the dressed holon $h_{i\sigma}$ is
a spinless fermion $h_{i}$ (bare holon) incorporated the spin
cloud $e^{-i\Phi_{i\sigma}}$ (magnetic flux), thus is a magnetic
dressing. In other words, the gauge invariant dressed holon
carries some spin messages, i.e., it shares its nontrivial spin
environment \cite{martins}. Although in common sense $h_{i\sigma}$
is not a real spinful fermion, it behaves like a spinful fermion.
The spirit of the charge-spin separation fermion-spin theory is
that the electron operator can be mapped as a product of the spin
operator and spinful fermion operator, this is very similar to the
case of the bosonization in one-dimensional interacting electron
systems, where the electron operators are mapped onto the boson
(electron density) representation, and then the recast Hamiltonian
is exactly solvable. In this charge-spin separation fermion-spin
representation, the low-energy behavior of the $t$-$t'$-$J$ model
(1) can be expressed as \cite{feng1},
\begin{eqnarray}\label{E6}
H&=&-t\sum_{i\hat{\eta}}(h_{i\uparrow}S^{+}_{i}
h^{\dagger}_{i+\hat{\eta}\uparrow}S^{-}_{i+\hat{\eta}}+
h_{i\downarrow}S^{-}_{i}h^{\dagger}_{i+\hat{\eta}\downarrow}
S^{+}_{i+\hat{\eta}})\nonumber\\
&+&t'\sum_{i\hat{\tau}}(h_{i\uparrow}S^{+}_{i}
h^{\dagger}_{i+\hat{\tau}\uparrow}S^{-}_{i+\hat{\tau}}+
h_{i\downarrow}S^{-}_{i}h^{\dagger}_{i+\hat{\tau}\downarrow}
S^{+}_{i+\hat{\tau}}) \nonumber \\
&-&\mu\sum_{i\sigma}h^{\dagger}_{i\sigma} h_{i\sigma}+J_{{\rm
eff}}\sum_{i\hat{\eta}}{\bf S}_{i}\cdot {\bf S}_{i+\hat{\eta}},
\end{eqnarray}
with $J_{{\rm eff}}=(1-x)^{2}J$, and $x=\langle
h^{\dagger}_{i\sigma}h_{i\sigma}\rangle=\langle h^{\dagger}_{i}
h_{i}\rangle$ is the hole doping concentration. As a consequence,
the kinetic energy terms in the $t$-$t'$-$J$ model have been
expressed as the dressed holon-spin interactions, which dominate
the essential physics of doped cuprates, while the magnetic energy
term is only to form an adequate dressed spin configuration
\cite{anderson2}. This reflects that even the kinetic energy terms
in the $t$-$t'$-$J$ Hamiltonian have strong Coulombic
contributions due to the restriction of no doubly occupancy of a
given site. This is why the interaction between the dressed holons
(dressed spins) can occur directly through the kinetic energy by
exchanging dressed spin (dressed holon) excitations
\cite{feng1,feng2}.

It has been shown that the superconducting state in cuprate
superconductors is characterized by the electron Cooper pairs,
forming superconducting quasiparticles \cite{tsuei}. Moreover, the
angle resolved photoemission spectroscopy measurements
\cite{shen1} show that in the real space the gap function and
pairing force have a range of one lattice spacing. These indicate
that the order parameter for the electron Cooper pair can be
expressed as,
\begin{eqnarray}\label{E7}
\Delta=\langle
C^{\dagger}_{i\uparrow}C^{\dagger}_{i+\hat{\eta}\downarrow}-
C^{\dagger}_{i\downarrow}C^{\dagger}_{i+\hat{\eta}\uparrow}\rangle
=\langle h_{i\uparrow}h_{i+\hat{\eta}\downarrow}S^{+}_{i}
S^{-}_{i+\hat{\eta}}-h_{i\downarrow}h_{i+\hat{\eta}\uparrow}
S^{-}_{i}S^{+}_{i+\hat{\eta}}\rangle .
\end{eqnarray}
In the doped regime without the antiferromagnetic long-range
order, the dressed spins form a disordered spin liquid state,
where the dressed spin correlation function $\langle
S^{+}_{i}S^{-}_{i+\hat{\eta}}\rangle=\langle S^{-}_{i}
S^{+}_{i+\hat{\eta}}\rangle$, then the order parameter for the
electron Cooper pair in Eq. (7) can be written as,
\begin{eqnarray}\label{E8}
\Delta=-\langle S^{+}_{i}S^{-}_{i+\hat{\eta}}\rangle \Delta_{h},
\end{eqnarray}
with the dressed holon pairing order parameter,
\begin{eqnarray}\label{E9}
\Delta_{h}=\langle h_{i+\hat{\eta}\downarrow}h_{i\uparrow}
-h_{i+\hat{\eta}\uparrow}h_{i\downarrow}\rangle,
\end{eqnarray}
which shows that the superconducting order parameter is related to
the dressed holon pairing amplitude, and is proportional to the
number of doped holes, and not to the number of electrons.
However, in the extreme low doped regime with the
antiferromagnetic long-range order, where the dressed spin
correlation function $\langle S^{+}_{i}S^{-}_{i+\hat{\eta}}
\rangle\neq\langle S^{-}_{i} S^{+}_{i+\hat{\eta}}\rangle$, then
the conduct is disrupted by the antiferromagnetic long-range
order, and therefore there is no mixing of superconductivity and
the antiferromagnetic long-range order \cite{bozovic}. In the case
without the antiferromagnetic long-range order, we
\cite{feng2,feng3} have discussed the kinetic energy driven
superconductivity within the $t$-$J$ model, and some qualitative
results are obtained. Following our previous discussions, we study
this issue within the $t$-$t'$-$J$ model for a quantitative
understanding of the physical properties of cuprates
superconductors.

The quantum spin operators obey the Pauli spin algebra, i.e., the
spin one-half raising and lowering operators $S^{+}_{i}$ and
$S^{-}_{i}$ behave as fermions on the same site and as bosons on
different sites, and therefore this problem can be discussed in
terms of the two-time spin Green's function within the Tyablikov
scheme \cite{tyablikov}. In the present case, we define the
dressed holon diagonal and off-diagonal Green's functions as,
\begin{eqnarray}
g(i-j,t-t')=-i\theta(t-t')\langle\{ h_{i\sigma}(t),
h^{\dagger}_{j\sigma}(t')\}\rangle =\langle\langle h_{i\sigma}(t);
h^{\dagger}_{j\sigma}(t') \rangle\rangle ,~~~~~\\
\Im (i-j,t-t')=-i\theta(t-t')\langle\{h_{i\downarrow}(t),
h_{j\uparrow}(t')\}\rangle =\langle\langle h_{i\downarrow}
(t);h_{j\uparrow}(t') \rangle\rangle ,~~~~~\\
\Im^{\dagger}(i-j,t-t')=-i\theta(t-t') \langle \{
h^{\dagger}_{i\uparrow}(t),h^{\dagger}_{j\downarrow}(t')\}\rangle
=\langle\langle h^{\dagger}_{i\uparrow}(t);
h^{\dagger}_{j\downarrow}(t')\rangle\rangle ,~~~~~
\end{eqnarray}
and the dressed spin Green's functions as,
\begin{eqnarray}
D(i-j,t-t')=-i\theta(t-t')\langle [S^{+}_{i}(t),S^{-}_{j}(t')]
\rangle=\langle\langle S^{+}_{i}(t);S^{-}_{j}(t')\rangle\rangle,
~~~~~\\
D_{z}(i-j,t-t')=-i\theta(t-t')\langle [S^{z}_{i}(t),S^{z}_{j}
(t')]\rangle=\langle\langle S^{z}_{i}(t);S^{z}_{j}(t')\rangle
\rangle, ~~~~~~
\end{eqnarray}
respectively, where $\langle \ldots \rangle$ is an average over
the ensemble. Within the mean-field approximation, the
$t$-$t'$-$J$ model can be decoupled as,
\begin{eqnarray}
H_{{\rm MFA}}&=&H_{t}+H_{J}+H_{0}, \\
H_{t}&=&\chi_{1}t\sum_{i\hat{\eta}\sigma}
h^{\dagger}_{i+\hat{\eta}\sigma} h_{i\sigma}
-\chi_{2}t'\sum_{i\hat{\tau}\sigma}
h^{\dagger}_{i+\hat{\tau}\sigma} h_{i\sigma}-\mu\sum_{i\sigma}
h^{\dagger}_{i\sigma}h_{i\sigma}, ~~~~~\\
H_{J}&=& {1\over 2}J_{{\rm eff}}\epsilon\sum_{i\hat{\eta}}
(S^{+}_{i}S^{-}_{i+\hat{\eta}}+S^{-}_{i}S^{+}_{i+\hat{\eta}})
+J_{{\rm eff}}\sum_{i\hat{\eta}}S^{z}_{i}S^{z}_{i+\hat{\eta}}
\nonumber \\
&-&t'\phi_{2}\sum_{i\hat{\tau}}(S^{+}_{i}S^{-}_{i+\hat{\tau}}+
S^{-}_{i}S^{+}_{i+\hat{\tau}}), \\
H_{0}&=& -2NZt\phi_{1}\chi_{1}+2NZt'\phi_{2}\chi_{2},
\end{eqnarray}
with $\epsilon=1+2t\phi_{1} /J_{{\rm eff}}$, $Z$ is the number of
the nearest neighbor or second-nearest neighbor sites, $N$ is the
number of sites, the dressed holon's particle-hole parameters
$\phi_{1}=\langle h^{\dagger}_{i\sigma}h_{i+\hat{\eta}\sigma}
\rangle$ and $\phi_{2}=\langle h^{\dagger}_{i\sigma}
h_{i+\hat{\tau}\sigma}\rangle$, and the spin correlation functions
$\chi_{1}=\langle S_{i}^{+} S_{i+\hat{\eta}}^{-}\rangle$ and
$\chi_{2}=\langle S_{i}^{+} S_{i+\hat{\tau}}^{-}\rangle$.
Therefore in the mean-field level, the dressed spin system is an
anisotropic away from the half-filling \cite{feng5}, and we have
defined the two dressed spin Green's function $D(i-j,t-t')$ in Eq.
(13) and $D_{z}(i-j,t-t')$ in Eq. (14) to describe the dressed
spin propagations. In the doped regime without the
antiferromagnetic long-range order, i.e., $\langle S^{z}_{i}
\rangle =0$, a mean-field theory of the $t$-$J$ model based on the
fermion-spin theory has been developed \cite{feng5} within the
Kondo-Yamaji decoupling scheme \cite{kondo}, which is a stage
one-step further than the Tyablikov's decoupling scheme. Following
their discussions \cite{feng5}, we can obtain the mean-field
dressed holon and spin Green's functions of the $t$-$t'$-$J$ model
as,
\begin{eqnarray}
g^{(0)}(k)&=&{1\over i\omega_{n}-\xi_{{\bf k}}}, \\
D^{(0)}(p)&=&{B_{{\bf p}}\over 2\omega_{{\bf p}}}\left ({1\over
ip_{m}-\omega_{{\bf p}}}-{1\over ip_{m}+\omega_{{\bf p}}}\right ),
\\
D^{(0)}_{z}(p)&=&{B_{z}({\bf p})\over 2\omega_{z}({\bf p})}\left (
{1\over  ip_{m}-\omega_{z}({\bf p})}-{1\over ip_{m}+\omega_{z}
({\bf p})} \right ),
\end{eqnarray}
respectively, where the four-vector notation $k=({\bf k},
i\omega_{n})$, $p=({\bf p},ip_{m})$, $B_{{\bf p}}=2\lambda_{1}
(A_{1}\gamma_{{\bf p}}-A_{2})-\lambda_{2}(2\chi^{z}_{2}
\gamma_{{\bf p}}'-\chi_{2})$, $B_{z}({\bf p})=\epsilon\chi_{1}
\lambda_{1}(\gamma_{{\bf p}}-1)-\chi_{2}\lambda_{2}(\gamma_{{\bf
p}}'-1)$, $\lambda_{1}= 2ZJ_{eff}$, $\lambda_{2} =4Z\phi_{2}t'$,
$\gamma_{{\bf k}}=(1/Z)\sum_{\hat{\eta}}e^{i{\bf k}\cdot
\hat{\eta}}$, $\gamma_{{\bf k}}'=(1/Z)\sum_{\hat{\tau}}e^{i{\bf k}
\cdot\hat{\tau}}$, $A_{1}=\epsilon\chi^{z}_{1}+\chi_{1}/2$, $A_{2}
=\chi^{z}_{1}+\epsilon\chi_{1}/2$, the spin correlation functions
$\chi^{z}_{1}=\langle S_{i}^{z} S_{i+\hat{\eta}}^{z}\rangle$ and
$\chi^{z}_{2}=\langle S_{i}^{z} S_{i+\hat{\tau}}^{z}\rangle$,  and
the mean-field dressed holon and spin excitation spectra are given
by,
\begin{eqnarray}
\xi_{{\bf k}}&=& Zt\chi_{1}\gamma_{{\bf k}}-Zt'\chi_{2}
\gamma_{{\bf k}}'-\mu, \\
\omega^{2}_{{\bf p}}&=& \lambda_{1}^{2}[(A_{4}-\alpha\epsilon
\chi^{z}_{1}\gamma_{{\bf p}}-{1\over 2Z}\alpha\epsilon\chi_{1})
(1-\epsilon\gamma_{{\bf p}})\nonumber \\
&+&{1\over 2}\epsilon(A_{3}-{1\over 2} \alpha\chi^{z}_{1}-\alpha
\chi_{1}\gamma_{{\bf p}})(\epsilon-\gamma_{{\bf p}})]
\nonumber \\
&+&\lambda_{2}^{2}[\alpha(\chi^{z}_{2}\gamma_{{\bf p}}'-{3\over
2Z}\chi_{2})\gamma_{{\bf p}}'+{1\over 2}(A_{5}-{1\over 2}
\alpha \chi^{z}_{2})]\nonumber \\
&+&\lambda_{1}\lambda_{2}[\alpha\chi^{z}_{1}(1-\epsilon
\gamma_{{\bf p}})\gamma_{{\bf p}}'+{1\over 2}\alpha(\chi_{1}
\gamma_{{\bf p}}'-C_{3})(\epsilon-\gamma_{{\bf p}})\nonumber\\
&+&\alpha \gamma_{{\bf p}}'(C^{z}_{3}-\epsilon \chi^{z}_{2}
\gamma_{{\bf p}})-{1\over 2}\alpha\epsilon(C_{3}- \chi_{2}
\gamma_{{\bf p}})], \\
\omega^{2}_{z}({\bf p})&=&\epsilon\lambda^{2}_{1}(\epsilon
A_{3}-{1\over Z}\alpha\chi_{1}-\alpha\chi_{1}\gamma_{{\bf p}})
(1-\gamma_{{\bf p}})+\lambda^{2}_{2}A_{5}(1-\gamma_{{\bf p}}')
\nonumber\\
&+&\lambda_{1}\lambda_{2}[\alpha\epsilon C_{3}(\gamma_{{\bf p}}+
\gamma_{{\bf p}}'-2)+\alpha\chi_{2}\gamma_{{\bf p}}(1-\gamma_{{\bf
p}}')],
\end{eqnarray}
with $A_{3}=\alpha C_{1}+(1-\alpha)/(2Z)$, $A_{4}=\alpha C^{z}_{1}
+(1-\alpha)/(4Z)$, $A_{5}=\alpha C_{2}+(1-\alpha)/(2Z)$, and the
spin correlation functions
$C_{1}=(1/Z^{2})\sum_{\hat{\eta},\hat{\eta'}}\langle
S_{i+\hat{\eta}}^{+}S_{i+\hat{\eta'}}^{-}\rangle$,
$C^{z}_{1}=(1/Z^{2})\sum_{\hat{\eta},\hat{\eta'}}\langle
S_{i+\hat{\eta}}^{z}S_{i+\hat{\eta'}}^{z}\rangle$,
$C_{2}=(1/Z^{2})\sum_{\hat{\tau},\hat{\tau'}}\langle
S_{i+\hat{\tau}}^{+}S_{i+\hat{\tau'}}^{-}\rangle$, and
$C_{3}=(1/Z)\sum_{\hat{\tau}}\langle S_{i+\hat{\eta}}^{+}
S_{i+\hat{\tau}}^{-}\rangle$, $C^{z}_{3}=(1/Z)
\sum_{\hat{\tau}}\langle S_{i+\hat{\eta}}^{z}
S_{i+\hat{\tau}}^{z}\rangle$. In order to satisfy the sum rule of
the correlation function $\langle S^{+}_{i}S^{-}_{i}\rangle=1/2$
in the case without the antiferromagnetic long-range order, the
important decoupling parameter $\alpha$ has been introduced in the
mean-field calculation \cite{feng5,kondo}, which can be regarded
as the vertex correction.

We \cite{feng2,feng3} have shown that the dressed holon-spin
coupling occurring in the kinetic energy terms in the $t$-$t'$-$J$
model (6) is quite strong. These interactions (kinetic energy
terms) can induce the dressed holon pairing state (then the
electron pairing state and superconductivity) by exchanging
dressed spin excitations in the higher power of the hole doping
concentration $x$. For discussion of superconductivity caused by
the strong dressed holon-spin interactions, we follow the
Eliashberg's strong coupling theory \cite{eliashberg}, and obtain
the self-consistent equations in terms of the equation of motion
method \cite{zubarev,feng1} that satisfied by the full dressed
holon diagonal and off-diagonal Green's functions as,
\begin{eqnarray}
g(k)&=&g^{(0)}(k)+g^{(0)}(k)[\Sigma^{(h)}_{1}(k)g(k)-
\Sigma^{(h)}_{2}(-k)\Im^{\dagger}(k)], \\
\Im^{\dagger}(k)&=&g^{(0)}(-k)[\Sigma^{(h)}_{1}(-k)
\Im^{\dagger}(-k)+\Sigma^{(h)}_{2}(-k)g(k)],
\end{eqnarray}
respectively, where the dressed holon self-energies are obtained
from the dressed spin bubble as \cite{feng1,feng2},
\begin{eqnarray}
\Sigma^{(h)}_{1}(k)&=&{1\over N^{2}}\sum_{{\bf p,p'}}
(Zt\gamma_{{\bf p+p'+k}}-Zt'\gamma_{{\bf p+p'+k}}')^{2}
{1\over \beta}\sum_{ip_{m}}g(p+k)\nonumber\\
&\times&{1\over\beta}\sum_{ip'_{m}}D^{(0)}(p')D^{(0)}(p'+p), \\
\Sigma^{(h)}_{2}(k)&=&{1\over N^{2}}\sum_{{\bf p,p'}}
(Zt\gamma_{{\bf p+p'+k}}-Zt'\gamma_{{\bf p+p'+k}}')^{2}
{1\over \beta}\sum_{ip_{m}}\Im (-p-k)\nonumber\\
&\times&{1\over\beta}\sum_{ip'_{m}} D^{(0)}(p')D^{(0)}(p'+p).
\end{eqnarray}
In the above calculations of the dressed holon self-energies, the
dressed spin part has been limited to the mean-field level, i.e.,
the full dressed spin Green's function $D(p)$ in Eqs. (27) and
(28) has been replaced as the mean-field dressed spin Green's
function (20), since the theoretical results of the normal-state
charge transport obtained at this level are consistent with the
experimental data \cite{feng1,feng6}.

In the self-consistent equations (25) and (26), the self-energy
function $\Sigma^{(h)}_{2}(k)$ describes the effective dressed
holon gap function, since both doping and temperature dependence
of the pairing force and dressed holon gap function have been
incorporated into $\Sigma^{(h)}_{2}(k)$. Furthermore, the
self-energy function $\Sigma^{(h)}_{1}(k)$ renormalizes the
mean-field dressed holon spectrum, and therefore it describes the
single particle (quasiparticle) coherence. Moreover, the
self-energy function $\Sigma^{(h)}_{2}(k)$ is an even function of
$i\omega_{n}$, while the other self-energy function
$\Sigma^{(h)}_{1}(k)$ is not. For the convenience, the self-energy
function $\Sigma^{(h)}_{1}(k)$ can be broken up into its symmetric
and antisymmetric parts as, $\Sigma^{(h)}_{1}(k)=\Sigma^{(h)}_{1e}
(k)+i\omega_{n}\Sigma^{(h)}_{1o}(k)$, then both $\Sigma^{(h)}_{1e}
(k)$ and $\Sigma^{(h)}_{1o}(k)$ are even functions of
$i\omega_{n}$. In this case, the charge carrier single particle
(quasiparticle) coherent weight can be defined as $Z^{-1}_{F}(k)
=1-\Sigma^{(h)}_{1o}(k)$, then the dressed holon diagonal and
off-diagonal Green's functions in Eqs. (25) and (26) can be
rewritten as,
\begin{eqnarray}
g(k)&=&{i\omega_{n} Z^{-1}_{F}(k)+\xi_{{\bf k}}+
\Sigma^{(h)}_{1e}(k) \over [i\omega_{n} Z^{-1}_{F}(k)]^{2}
-[\xi_{{\bf k}}+\Sigma^{(h)}_{1e}(k)]^{2}-
[\Sigma^{(h)}_{2}(k)]^{2}} ,\\
\Im^{\dagger}(k)&=&-{\Sigma^{(h)}_{2}(k)\over [i\omega_{n}
Z^{-1}_{F}(k)]^{2}-[\xi_{{\bf k}}+\Sigma^{(h)}_{1e} (k)]^{2}-
[\Sigma^{(h)}_{2}(k)]^{2}}.
\end{eqnarray}
As in the conventional superconductor \cite{eliashberg}, the
retarded function ${\rm Re}\Sigma^{(h)}_{1e} (k)$ is a constant,
independent of (${\bf k},\omega$), and therefore it just
renormalizes the chemical potential. In this case, it can be
dropped. Furthermore, we only study the static limit of the
effective dressed holon gap function and single particle coherent
weight, i.e., $\Sigma^{(h)}_{2}(k)= \bar{\Delta}_{h}({\bf k})$,
and $Z^{-1}_{F}({\bf k})=1- \Sigma^{(h)}_{1o}({\bf k})$, then the
dressed holon diagonal and off-diagonal Green's functions in Eqs.
(29) and (30) can be expressed explicitly as,
\begin{eqnarray}
g(k)&=&{1\over 2}\left (1+{\bar{\xi_{{\bf k}}}\over E_{{\bf k}}}
\right ){Z_{F}({\bf k})\over i\omega_{n}-E_{{\bf k}}}+{1\over 2}
\left (1- {\bar{\xi_{{\bf k}}}\over E_{{\bf k}}}
\right ){Z_{F}({\bf k})\over i\omega_{n}+E_{{\bf k}}},~~~~~ \\
\Im^{\dagger}(k)&=&-{Z_{F}({\bf k})\bar{\Delta}_{hZ}({\bf k})\over
2E_{{\bf k}}}\left ( {1\over i\omega_{n}-E_{{\bf k}}}- {1\over
i\omega_{n}+ E_{{\bf k}}}\right ),
\end{eqnarray}
with $\bar{\xi_{{\bf k}}}=Z_{F}({\bf k})\xi_{{\bf k}}$,
$\bar{\Delta}_{hZ}({\bf k})=Z_{F}({\bf k})\bar{\Delta}_{h}({\bf
k})$, and the dressed holon quasiparticle spectrum $E_{{\bf k}}=
\sqrt {\bar{\xi^{2}_{{\bf k}}}+\mid\bar{\Delta}_{hZ}({\bf k})
\mid^{2}}$. Although $Z_{F}({\bf k})$ still is a function of ${\bf
k}$, the wave vector dependence is unimportant, since everything
happens near the electron Fermi surface. Therefore we need to
estimate the special wave vector ${\bf k}_{0}$ that guarantees
$Z_{F}=Z_{F} ({\bf k}_{0})$ near the electron Fermi surface. In
the present charge-spin separation fermion-spin framework
\cite{feng5}, the electron diagonal Green's function $G(i-j,t-t')=
\langle\langle C_{i\sigma} (t);C^{\dagger}_{j\sigma}(t') \rangle
\rangle$ is a convolution of the dressed spin Green's function
$D(p)$ and dressed holon diagonal Green's function $g(k)$, and can
be evaluated  as \cite{feng5},
\begin{eqnarray}
G(k)&=&{1\over N}\sum_{{\bf p}}\int^{\infty}_{-\infty}{d\omega'
\over 2\pi}\int^{\infty}_{-\infty}{d\omega''\over 2\pi}A_{s}({\bf
p},\omega') A_{h}({\bf
p+k},\omega'')\nonumber \\
&\times& {n_{F}(\omega'')+n_{B}(\omega')\over i\omega_{n}
+\omega''-\omega'},
\end{eqnarray}
where the dressed spin spectral function $A_{s}({\bf p},\omega)
=-2{\rm Im}D({\bf p},\omega)$, the dressed holon spectral function
$A_{h}({\bf k},\omega) =-2{\rm Im}g({\bf k},\omega)$, and $n_{B}
(\omega)$ and $n_{F}(\omega)$ are the boson and fermion
distribution functions, respectively. This convolution of the
dressed spin Green's function and diagonal dressed holon Green's
function reflects the charge-spin recombination \cite{anderson2}.
With the help of this electron diagonal Green's function, the
electron spectral function $A({\bf k},\omega) =-2{\rm Im} G({\bf
k},\omega)$ is obtained as,
\begin{eqnarray}
A({\bf k},\omega)&=&{1\over N}\sum_{{\bf p}}
\int^{\infty}_{-\infty}{d\omega' \over 2\pi}A_{s}({\bf p},
\omega')A_{h}({\bf p+k},\omega'+\omega) \nonumber \\
&\times& [n_{F}(\omega'+\omega)+n_{B}(\omega')].
\end{eqnarray}
This electron spectral function has been used to extract the
electron momentum distribution (then the electron Fermi surface)
as \cite{feng5},
\begin{eqnarray}
n_{{\bf k}}&=&\int^{\infty}_{-\infty}{d\omega\over 2\pi}A({\bf k},
\omega)n_{F}(\omega) \nonumber \\
&=& {1\over 2}-{1\over N}\sum_{{\bf p}}n_{s}({\bf p})
\int^{\infty}_{-\infty}{d\omega\over 2\pi}A_{h}({\bf p+k},
\omega)n_{F}(\omega), ~~~~~
\end{eqnarray}
where
\begin{eqnarray}
n_{s}({\bf p})=\int^{\infty}_{-\infty}{d\omega\over 2\pi} A_{s}
({\bf p},\omega)n_{B}(\omega),
\end{eqnarray}
is the dressed spin momentum distribution. In the present case,
this electron momentum distribution can be evaluated explicitly in
terms of the mean-field dressed spin Green's function (20) and
dressed holon diagonal Green's function (31) as,
\begin{eqnarray}
n_{{\bf k}}={1\over 2}-{1\over 2N}\sum_{{\bf p}}n^{(0)}_{s}({\bf
p})Z_{F}({\bf p-k})\left (1- {\bar{\xi}_{{\bf p-k}}\over E_{{\bf
p-k}}}{\rm tanh} [{1\over 2}\beta E_{{\bf p-k}}]\right ),
\end{eqnarray}
with $n^{(0)}_{s}({\bf p})=B_{{\bf p}}{\rm coth}(\beta\omega_{{\bf
p}}/2)/(2\omega_{{\bf p}})$. Since the dressed spins center around
$[\pm\pi,\pm\pi]$ in the Brillouin zone in the mean-field level
\cite{feng5}, therefore the above electron momentum distribution
can be approximately reduced as $n_{{\bf k}}\approx
1/2-\rho^{(0)}_{s} Z_{F}({\bf k_{A}-k})[1- \bar{\xi}_{{\bf
k_{A}-k}}{\rm tanh}(\beta E_{{\bf k_{A}-k}}/2)/ E_{{\bf
k_{A}-k}}]/2$, with ${\bf k_{A}}=[\pi,\pi]$, and
$\rho^{(0)}_{s}=(1/N)\sum_{{\bf p}= (\pm\pi, \pm\pi)}n^{(0)}_{s}
({\bf p})$. It has been shown from the angle resolved
photoemission spectroscopy experiments \cite{shen2} that the
electron Fermi surface is small pockets around $[\pi/2,\pi/2]$ at
small doping, and becomes a large electron Fermi surface at large
doping. In this case, the Fermi wave vector can be estimated
\cite{feng5} at ${\bf k_{F}}=[(1-x)\pi/2,(1-x)\pi/2]$, and is
evolution with doping. Now we obtain the wave vector ${\bf k}_{0}
\approx {\bf k_{A}}-{\bf k_{F}}$ that guarantees $Z_{F}({\bf
k}_{0})$ near the electron Fermi surface, and we only need to
calculate $Z_{F}=Z_{F}({\bf k}_{0})$ as mentioned above. Since the
charge-spin recombination from the convolution of the dressed spin
Green's function and dressed holon diagonal Green's function leads
to form the electron Fermi surface \cite{anderson2}, then the
dressed holon single particle coherence $Z_{F}$ appearing in the
electron momentum distribution also reflects the electron single
particle coherence.

Experimentally, some results seem consistent with an s-wave
pairing \cite{chaudhari}, while other measurements gave the
evidence in favor of the d-wave pairing \cite{martindale,tsuei}.
These reflect a fact that the d-wave gap function $\propto {\rm
k}^{2}_{x}-{\rm k}^{2}_{y}$ belongs to the same representation
$\Gamma_{1}$ of the orthorhombic crystal group as does s-wave gap
function $\propto {\rm k}^{2}_{x}+{\rm k}^{2}_{y}$. For
understanding of these experimental results, we consider both
s-wave and d-wave cases, i.e., $\bar{\Delta}^{(s)}_{hZ}({\bf k})=
\bar{\Delta}^{(s)}_{hZ}\gamma^{(s)}_{{\bf k}}$, with
$\gamma^{(s)}_{{\bf k}}=\gamma_{{\bf k}}=({\rm cos}k_{x}+{\rm cos}
k_{y})/2$, for the s-wave pairing, and $\bar{\Delta}^{(d)}_{hZ}
({\bf k})=\bar{\Delta}^{(d)}_{hZ}\gamma^{(d)}_{{\bf k}}$, with
$\gamma^{(d)}_{{\bf k}}=({\rm cos} k_{x}-{\rm cos}k_{y})/2$, for
the d-wave pairing, respectively. In this case, the dressed holon
effective gap parameter and single particle coherent weight in
Eqs. (27) and (28) satisfy following two equations
\cite{feng2,feng3},
\begin{eqnarray}
1&=&{1\over N^{3}}\sum_{{\bf k,q,p}}(Zt\gamma_{{\bf k+q}}-Zt'
\gamma_{{\bf k+q}}')^{2}\gamma^{(a)}_{{\bf k-p+q}}
\gamma^{(a)}_{{\bf k}}{Z^{2}_{F}\over E_{{\bf k}}}{B_{{\bf
q}}B_{{\bf p}}\over\omega_{{\bf q}} \omega_{{\bf p}}}\nonumber\\
&\times& \left({F^{(1)}_{1}({\bf k,q,p})\over (\omega_{{\bf p}}-
\omega_{{\bf q}})^{2}-E^{2}_{{\bf k}}}-{F^{(2)}_{1}({\bf k,q,p})
\over (\omega_{{\bf p}}+\omega_{{\bf q}})^{2}- E^{2}_{{\bf k}}}
\right ) ,\\
Z^{-1}_{F}&=&1+{1\over N^{2}}\sum_{{\bf q,p}}(Zt\gamma_{{\bf
p+k_{0}}}-Zt'\gamma_{{\bf p+k_{0}}}')^{2}Z_{F}{B_{{\bf q}}
B_{{\bf p}}\over 4\omega_{{\bf q}}\omega_{{\bf p}}}\nonumber\\
&\times& \left({F^{(1)}_{2}({\bf q,p})\over (\omega_{{\bf p}}-
\omega_{{\bf q}}-E_{{\bf p-q+k_{0}}})^{2}}+{F^{(2)}_{2}({\bf
q,p})\over (\omega_{{\bf p}}-\omega_{{\bf q}}+E_{{\bf
p-q+k_{0}}})^{2}} \right. \nonumber \\
&+& \left . {F^{(3)}_{2}({\bf q,p})\over (\omega_{{\bf p}}+
\omega_{{\bf q}}-E_{{\bf p-q+k_{0}}})^{2}}+{F^{(4)}_{2}({\bf
q,p})\over (\omega_{{\bf p}}+\omega_{{\bf q}}+E_{{\bf
p-q+k_{0}}})^{2}} \right ) ,~~~~~
\end{eqnarray}
respectively, where $a={\rm s,d}$, and
\begin{eqnarray}
F^{(1)}_{1}({\bf k,q,p})&=&(\omega_{{\bf p}}-\omega_{{\bf q}})
[n_{B}(\omega_{{\bf q}})-n_{B}(\omega_{{\bf p}})][1-2 n_{F}
(E_{{\bf k}})] \nonumber \\
&+&E_{{\bf k}}[n_{B}(\omega_{{\bf p}})n_{B} (-\omega_{{\bf q}})
+n_{B}(\omega_{{\bf q}})n_{B}(-\omega_{{\bf p}})], \\
F^{(2)}_{1}({\bf k,q,p})&=&(\omega_{{\bf p }}+\omega_{{\bf q}})
[n_{B}(-\omega_{{\bf p}})-n_{B}(\omega_{{\bf q}})][1-2 n_{F}
(E_{{\bf k}})] \nonumber \\
&+&E_{{\bf k}}[n_{B}(\omega_{{\bf p}}) n_{B} (\omega_{{\bf q}})+
n_{B}(-\omega_{{\bf p}})n_{B}(- \omega_{{\bf q} })], \\
F^{(1)}_{2}({\bf q,p})&=&n_{F}(E_{{\bf p-q+k_{0}}})[n_{B}
(\omega_{{\bf q}})-n_{B}(\omega_{{\bf p}})] \nonumber \\
&-& n_{B}(\omega_{{\bf p}})n_{B}(-\omega_{{\bf q}}), ~~~~~~\\
F^{(2)}_{2} ({\bf q,p})&=&n_{F}(E_{{\bf p-q+k_{0}}})[n_{B}
(\omega_{{\bf p}})-n_{B} (\omega_{{\bf q}})]\nonumber \\
&-&n_{B}(\omega_{{\bf q}}) n_{B} (-\omega_{{\bf p}}), ~~~~~~\\
F^{(3)}_{2}({\bf q,p})&=& n_{F}(E_{{\bf p-q+k_{0}}}) [n_{B}
(\omega_{{\bf q}})-n_{B}(-\omega_{{\bf p}})] \nonumber \\
&+&n_{B}(\omega_{{\bf p}})n_{B}(\omega_{{\bf q}}), ~~~~~~\\
F^{(4)}_{2}({\bf q,p})&=&n_{F}(E_{{\bf p-q+k_{0}}})
[n_{B}(-\omega_{{\bf q}})-n_{B} (\omega_{{\bf p}})] \nonumber \\
&+& n_{B}(-\omega_{{\bf p}})n_{B}(-\omega_{{\bf q}}). ~~~~~~
\end{eqnarray}
These two equations (38) and (39) must be solved simultaneously
with other self-consistent equations \cite{feng2},
\begin{eqnarray}
\phi_{1}&=&{1\over 2N}\sum_{{\bf k}}\gamma_{{\bf k}}Z_{F}\left
(1-{\bar{\xi_{{\bf k}}}\over E_{{\bf k}}}{\rm th}
[{1\over 2}\beta E_{{\bf k}}]\right ),\\
\phi_{2}&=&{1\over 2N}\sum_{{\bf k}}\gamma_{{\bf k}}'Z_{F}\left
(1-{\bar{\xi_{{\bf k}}}\over E_{{\bf k}}}{\rm th}
[{1\over 2}\beta E_{{\bf k}}]\right ),\\
\delta &=& {1\over 2N}\sum_{{\bf k}}Z_{F}\left (1-{\bar{\xi_{{\bf
k}}} \over E_{{\bf k}}}{\rm th}[{1\over 2}\beta E_{{\bf k}}]
\right ),\\
\chi_{1}&=&{1\over N}\sum_{{\bf k}}\gamma_{{\bf k}} {B_{{\bf
k}}\over 2\omega_{{\bf k}}}{\rm coth}
[{1\over 2}\beta\omega_{{\bf k}}],\\
\chi_{2}&=&{1\over N}\sum_{{\bf k}}\gamma_{{\bf k}}'{B_{{\bf
k}}\over 2\omega_{{\bf k}}}{\rm coth}
[{1\over 2}\beta\omega_{{\bf k}}],\\
C_{1}&=&{1\over N}\sum_{{\bf k}}\gamma^{2}_{{\bf k}} {B_{{\bf
k}}\over 2\omega_{{\bf k}}}{\rm coth}
[{1\over 2}\beta\omega_{{\bf k}}],\\
C_{2}&=&{1\over N}\sum_{{\bf k}}\gamma'^{2}_{{\bf k}} {B_{{\bf
k}}\over 2\omega_{{\bf k}}}{\rm coth}  [{1\over 2}
\beta\omega_{{\bf k}}], \\
C_{3}&=&{1\over N}\sum_{{\bf k}}\gamma_{{\bf k}}\gamma_{{\bf k}}'
{B_{{\bf k}}\over 2\omega_{{\bf k}}}{\rm coth} [{1\over 2}
\beta\omega_{{\bf k}}],
\end{eqnarray}
\begin{eqnarray}
{1\over 2} &=&{1\over N}\sum_{{\bf k}}{B_{{\bf k}} \over
2\omega_{{\bf k}}}{\rm coth} [{1\over 2}\beta\omega_{{\bf k}}],\\
\chi^{z}_{1}&=&{1\over N}\sum_{{\bf k}}\gamma_{{\bf k}}
{B_{z}({\bf k})\over 2\omega_{z}({\bf k})}{\rm coth}
[{1\over 2}\beta\omega_{z}({\bf k})],\\
\chi^{z}_{2}&=&{1\over N}\sum_{{\bf k}}\gamma_{{\bf k}}'
{B_{z}({\bf k})\over 2\omega_{z}({\bf k})}{\rm coth}
[{1\over 2}\beta\omega_{z}({\bf k})],\\
C^{z}_{1}&=&{1\over N}\sum_{{\bf k}}\gamma^{2}_{{\bf k}}
{B_{z}({\bf k})\over 2\omega_{z}({\bf k})}{\rm coth} [{1\over
2}\beta\omega_{z}({\bf k})],\\
C^{z}_{3}&=&{1\over N}\sum_{{\bf k}}\gamma_{{\bf k}}\gamma_{{\bf
k}}'{B_{z}({\bf k})\over 2\omega_{z}({\bf k})}{\rm coth} [{1\over
2}\beta\omega_{z}({\bf k})],
\end{eqnarray}
then all the order parameters, decoupling parameter $\alpha$, and
chemical potential $\mu$ are determined by the self-consistent
calculation \cite{feng2}. With above discussions, we now can
obtain the dressed holon pair gap function in terms of the
off-diagonal Green's function (32) as,
\begin{eqnarray}
\Delta^{(a)}_{h}({\bf k})=-{1\over \beta}\sum_{i\omega_{n}}
\Im^{\dagger}({\bf k},i\omega_{n})={1\over 2}Z_{F}
{\bar{\Delta}^{(a)}_{hZ}({\bf k})\over E_{{\bf k}}}{\rm tanh}
[{1\over 2}\beta E_{{\bf k}}],
\end{eqnarray}
then the dressed holon pair order parameter in Eq. (9) can be
evaluated explicitly from this dressed holon pair gap function
(59) as,
\begin{eqnarray}
\Delta^{(a)}_{h}={2\over N}\sum_{{\bf k}} [\gamma^{(a)}_{{\bf k}}
]^{2} {Z_{F}\bar{\Delta}^{(a)}_{hZ}\over E_{{\bf k}}}{\rm tanh}
[{1\over 2}\beta E_{{\bf k}}].
\end{eqnarray}
In the charge-spin separation fermion-spin theory, we \cite{feng2}
have shown that the dressed holon pairing state originating from
the kinetic energy terms by exchanging dressed spin excitations
also leads to form the electron Cooper pairing state, and the
superconducting gap function is obtained from the electron
off-diagonal Green's function $\Gamma^{\dagger}(i-j,t-t')=\langle
\langle C^{\dagger}_{i\uparrow}(t);C^{\dagger}_{j\downarrow}(t')
\rangle\rangle$, which is a convolution of the dressed spin
Green's function and dressed holon off-diagonal Green's function
and reflects the charge-spin recombination \cite{anderson2}. In
the present case, it can be evaluated in terms of the mean-field
dressed spin Green's function (20) and dressed holon off-diagonal
Green's function (32) as,
\begin{eqnarray}
\Gamma^{\dagger}(k)&=&{1\over N}\sum_{{\bf p}}
{Z_{F}\bar{\Delta}^{(a)}_{hZ}({\bf p-k})\over E_{{\bf p-k}}}
{B_{{\bf p}} \over 2\omega_{{\bf p}}}\left ({(\omega_{{\bf p}}
+E_{{\bf p-k}})[n_{B}(\omega_{{\bf p}})+n_{F}(-E_{{\bf p-k}})]
\over (i\omega_{n})^{2}-(\omega_{{\bf p}}+E_{{\bf p-k}})^{2}}
\right . \nonumber \\
&-& \left . {(\omega_{{\bf p}}-E_{{\bf p-k}})[n_{B}(\omega_{{\bf
p}})+n_{F}(E_{{\bf p-k}})]\over (i\omega_{n})^{2}-(\omega_{{\bf
p}}-E_{{\bf p-k}})^{2}} \right ).
\end{eqnarray}
then the superconducting gap function is obtained from this
electron off-diagonal Green's function as,
\begin{eqnarray}
\Delta^{(a)}({\bf k})=-{1\over N}\sum_{{\bf p}}
{Z_{F}\bar{\Delta}^{(a)}_{Zh}({\bf p-k})\over 2E_{{\bf p-k}}}{\rm
tanh}[{1\over 2}\beta E_{{\bf p-k}}]{B_{{\bf p}}\over
2\omega_{{\bf p}}} {\rm coth}[{1\over 2}\beta\omega_{{\bf p}}],
\end{eqnarray}
which shows that the symmetry of the electron Cooper pair is
essentially determined by the symmetry of the dressed holon pair.
From this superconducting gap function (62), the superconducting
gap parameter in Eq. (8) is obtained in terms of Eqs. (60) and
(49) as $\Delta^{(a)}=-\chi_{1} \Delta^{(a)}_{h}$. However, in
contrast with the conventional superconductors, the dressed holon
(then electron) pairing interaction in cuprate superconductors is
also doping dependent, therefore the experimental observed
superconducting gap parameter in cuprate superconductors should be
the effective superconducting gap parameter $\bar{\Delta}^{(a)}
\sim -\chi_{1}\bar {\Delta}^{(a)}_{h}$, which measures the
strength of the binding of electrons into electron Cooper pairs.
In Fig. 1, we plot the effective dressed holon pairing (a) and
effective superconducting (b) gap parameters in the s-wave
symmetry (solid line) and d-wave symmetry (dashed line) as a
function of the hole doping concentration $x$ at $T=0.002J$ for
$t/J=2.5$ and $t'/J=0.3$. For comparison, the experimental result
\cite{wen} of the upper critical field as a function of the hole
doping concentration is also shown in Fig. 1(b). In a given doping
concentration, the upper critical field is defined as the critical
field that destroys the superconducting-state at the zero
temperature, therefore the upper critical field also measures the
strength of the binding of electrons into Cooper pairs like the
effective superconducting gap parameter \cite{wen}. In other
words, both effective superconducting gap parameter and upper
critical field have a similar doping dependence \cite{wen}. In
this sense, our result is in good agreement with the experimental
data \cite{wen}. In particular, the value of $\bar{\Delta}^{(d)}$
increases with increasing doping in the underdoped regime, and
reaches a maximum in the optimal doping $x_{{\rm opt}}\approx
0.15$, then decreases in the overdoped regime. Since the effective
dressed holon pairing gap parameter measures the strength of the
binding of dressed holons into dressed holon pairs, then our
results also show that although the superconductivity is driven by
the kinetic energy by exchanging dressed spin excitations, the
strength of the binding of electrons into electron Cooper pairs is
still suppressed by the short-range antiferromagnetic fluctuation.

\begin{figure}[t]
\begin{center}
\begin{minipage}[h]{125mm}
\epsfig{file=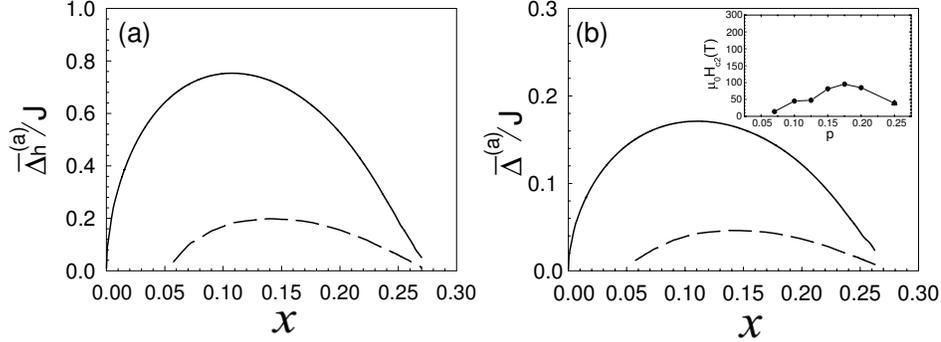, width=125mm}
\end{minipage}
\caption{The effective dressed holon pairing (a) and effective
superconducting (b) gap parameters in the s-wave symmetry (solid
line) and d-wave symmetry (dashed line) as a function of the hole
doping concentration in $T=0.002J$ for $t/J=2.5$ and $t'/t=0.3$.
Inset: the experimental result of the upper critical field as a
function of the hole doping concentration taken from Ref.
\protect\cite{wen}.} \label{F1}
\end{center}
\end{figure}

Now we turn to discuss the superconducting transition temperature.
The present result in Eq. (62) also shows that the superconducting
transition temperature $T^{(a)}_{c}$ occurring in the case of the
superconducting gap parameter $\Delta^{(a)}=0$ is identical to the
dressed holon pair transition temperature occurring in the case of
the effective holon pairing gap parameter $\bar{\Delta}^{(a)}_{hZ}
=0$. In this case, we have performed a calculation for the doping
dependence of the superconducting transition temperature, and the
result of the superconducting transition temperature $T^{(a)}_{c}$
as a function of the hole doping concentration $x$ in the s-wave
symmetry (solid line) and d-wave symmetry (dashed line) for
$t/J=2.5$ and $t'/J=0.3$ is plotted in Fig. 2 in comparison with
the experimental result \cite{tallon} (inset). Our result shows
that for the s-wave symmetry, the maximal superconducting
transition temperature T$^{(s)}_{c}$ occurs around a particular
doping concentration $x\approx 0.10$, and then decreases in both
lower doped and higher doped regimes. However, for the d-wave
symmetry, the maximal superconducting transition temperature
T$^{(d)}_{c}$ occurs around the optimal doping concentration
$x_{{\rm opt}}\approx 0.15$, and then decreases in both underdoped
and overdoped regimes. Although the superconducting pairing
symmetry is doping dependent, the superconducting state has the
d-wave symmetry in a wide range of doping, in good agreement with
the experiments \cite{yeh,biswas,tsuei1}. Furthermore,
T$^{(d)}_{c}$ in the underdoped regime (T$^{(s)}_{c}$ in the lower
doped regime) is proportional to the hole doping concentration
$x$, and therefore T$^{(d)}_{c}$ in the underdoped regime
(T$^{(s)}_{c}$ in the lower doped regime) is set by the hole
doping concentration \cite{uemura}. This reflects that the density
of the dressed holons directly determines the superfluid density
in the underdoped regime for the d-wave case (the lower doped
regime for the s-wave case). Using an reasonably estimative value
of $J\sim 800$K to 1200K in doped cuprates \cite{shamoto}, the
superconducting transition temperature in the optimal doping is
T$^{(d)}_{c} \approx 0.22J \approx 176{\rm K}\sim 264{\rm K}$, in
semiquantitative agreement with the experimental data
\cite{tallon,tsuei1,uemura}. In comparison with the previous
result based on the $t$-$J$ model \cite{feng3}, our present result
of the $t$-$t'$-$J$ model also shows that the effect of the
additional second neighbor hopping $t'$ is to enhance the d-wave
superconducting pairing correlation, and suppress the s-wave
superconducting pairing correlation.

\begin{figure}[t]
\begin{center}
\begin{minipage}[h]{85mm}
\epsfig{file=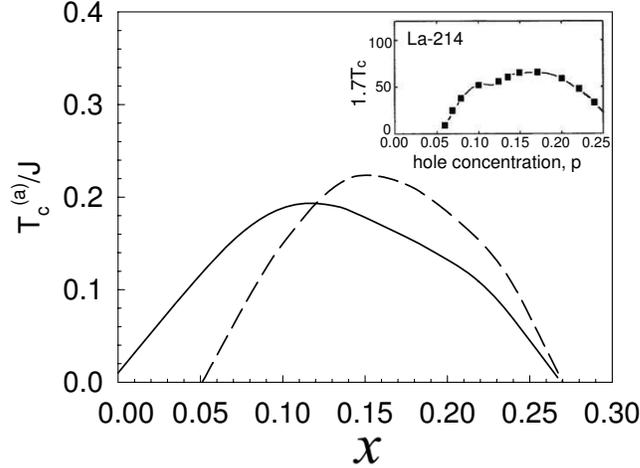, width=85mm}
\end{minipage}
\caption{The superconducting transition temperature as a function
of the hole doping concentration in the s-wave symmetry (solid
line) and d-wave symmetry (dashed line) for $t/J=2.5$ and
$t'/t=0.3$. Inset: the experimental result taken from Ref.
\protect\cite{tallon}.} \label{F2}
\end{center}
\end{figure}

Since cuprates superconductors are highly anisotropic materials,
therefore the electron spectral function $A({\bf k}, \omega)$ is
dependent on the in-plane momentum \cite{shen2}. Although the
electron spectral function in doped cuprates obtained from the
angle resolved photoemission spectroscopy is very broad in the
normal-state, indicating that there are no quasiparticles.
However, in the superconducting-state, the full energy dispersion
of quasiparticles has been observed \cite{ding2}. According to a
comparison of the density of states as measured by scanning
tunnelling microscopy \cite{dewilde} and angle resolved
photoemission spectroscopy spectral function \cite{ding,shen2} at
$[\pi,0]$ point on identical samples, it has been shown that the
most contributions of the angle-integrated spectral function come
from $[\pi,0]$ point. In addition, the d-wave gap, and therefore
the electron pairing energy scale, is maximized at $[\pi,0]$
point. Although the sharp superconducting quasiparticle peak at
$[\pi,0]$ point in cuprate superconductors has been widely
studied, the orgin and its implications are still under debate
\cite{ding2}. Within the present theoretical framework of
superconductivity, we have discussed this issue, and the result of
the electron spectral function (34) with the d-wave symmetry at
$[\pi,0]$ point in the optimal doping $x_{{\rm opt}}=0.15$ and
temperature $T=0.002J$ for $t/J=2.5$ and $t'/J=0.3$ is plotted in
Fig. 3 in comparison with the experimental result \cite{ding}
(inset). Our result shows that the position of the peak of the
electron spectral function in $[\pi,0]$ point is located at
$\omega_{{\rm peak}}\approx 0.4J \approx 0.028$eV$\sim 0.04$eV,
which is quantitatively consistent with the $\omega_{{\rm peak}}
\approx 0.03$eV observed \cite{ding} in the cuprate superconductor
Bi$_{2}$Sr$_{2}$CaCu$_{2}$O$_{8+x}$. Furthermore, we have
discussed the temperature dependence of the electron spectral
function and overall quasiparticle dispersion, and these and
related theoretical results will be presented elsewhere.

\begin{figure}[t]
\begin{center}
\begin{minipage}[h]{85mm}
\epsfig{file=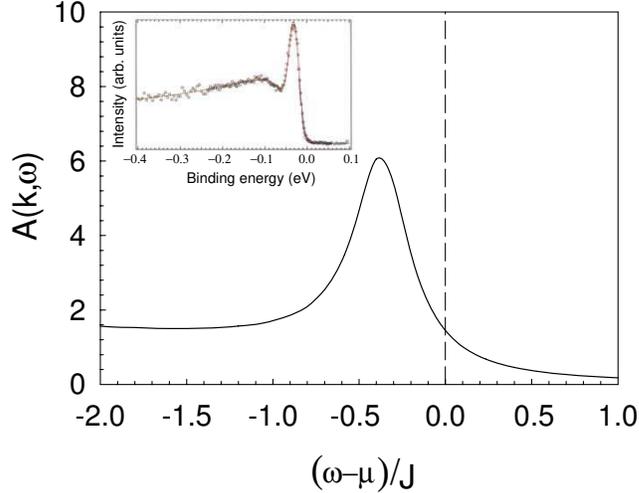, width=85mm}
\end{minipage}
\caption{The electron spectral function with the d-wave symmetry
at $[\pi,0]$ point in $x_{{\rm opt}}=0.15$ and $T=0.002J$ for
$t/J=2.5$ and $t'/J=0.3$.  Inset: the experimental result taken
from Ref. \protect\cite{ding}.} \label{F3}
\end{center}
\end{figure}

\begin{figure}[t]
\begin{center}
\begin{minipage}[h]{85mm}
\epsfig{file=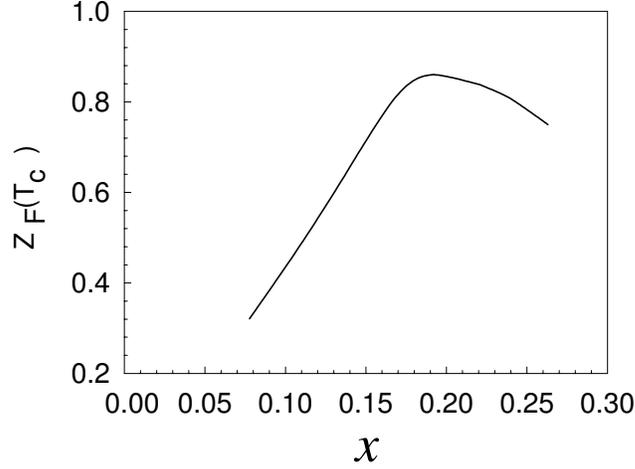, width=85mm}
\end{minipage}
\caption{The single particle coherent weight $Z_{F}(T_{c})$ as a
function of the hole doping concentration for $t/J=2.5$ and
$t'/t=0.3$.} \label{F4}
\end{center}
\end{figure}

The essential physics of superconductivity in the present
$t$-$t'$-$J$ model is the same as that in the $t$-$J$ model
\cite{feng2,feng3}. The antisymmetric part of the self-energy
function $\Sigma^{(h)}_{1o}({\bf k})$ (then $Z_{F}$) describes the
single particle (quasiparticle) coherence, and therefore $Z_{F}$
is closely related to the quasiparticle density, while the
self-energy function $\Sigma^{(h)}_{2}({\bf k})$ describes the
effective dressed holon pairing gap function. In particular, both
$Z_{F}$ and $\Sigma^{(h)}_{2}({\bf k})$ are doping and temperature
dependent. Since the superconducting-order is established through
an emerging quasiparticle \cite{ding}, therefore the
superconducting-order is controlled by both gap function and
quasiparticle coherence, and is reflected explicitly in the
self-consistent equations (38) and (39). To show this point
clearly, we plot the single particle (quasiparticle) coherent
weight $Z_{F}(T_{c})$ as a function of the hole doping
concentration $x$ for $t/J=2.5$ and $t'/t=0.3$ in Fig. 4. As seen
from Fig. 4, the doping dependent behavior of the single particle
coherent weight resembles that of the superfluid density in
cuprate superconductors, i.e., $Z_{F}(T_{c})$ grows linearly with
the hole doping concentration in the underdoped and optimally
doped regimes, and then decreases with increasing doping in the
overdoped regime, which leads to that the superconducting
transition temperature reaches a maximum in the optimal doping,
and then decreases in both underdoped and overdoped regimes. Since
the effective superconducting gap function
$\bar{\Delta}^{(s)}({\bf k})=\bar{\Delta}^{(s)}({\rm cos}k_{x}
+{\rm cos}k_{y})/2$ for the s-wave symmetry or
$\bar{\Delta}^{(d)}({\bf k})=\bar{\Delta}^{(d)}({\rm cos}k_{x}
-{\rm cos}k_{y})/2$ for the d-wave case is dependent on the
momentum and the most contributions of the electronic states come
from $[\pi,0]$ point \cite{ding}, therefore although the value of
the effective superconducting gap parameter $\bar{\Delta}^{(s)}$
(then the ratio $\bar{\Delta}^{(s)}$/T$^{(s)}_{c}$) for the s-wave
symmetry is larger than these $\bar{\Delta}^{(d)}$ (then the ratio
$\bar{\Delta}^{(d)}$/T$^{(d)}_{c}$) in the d-wave case, the system
has the superconducting transition temperature T$^{(d)}_{c}$ with
the d-wave symmetry in a wide range of doping. Although the
superconducting state is characterized by the charge-spin
recombination, the superconducting transition temperature is
determined by the dressed holon pairing transition temperature.
This is why the superfluid density in the underdoped regime
vanishes more or less linearly with the hole doping concentration
\cite{uemura}, and the cuprate superconductors are the gossamer
superconductors. Moreover, the dressed holon pairs condense with
the d-wave symmetry in a wide range of doping in the present
framework of superconductivity, then the electron Cooper pairs
originating from the dressed holon pairing state are due to the
charge-spin recombination, and their condensation automatically
gives the electron quasiparticle character. Our results also show
that the superconducting-state of cuprate superconductors is the
conventional Bardeen-Cooper-Schrieffer like, so that some of the
basic Bardeen-Cooper-Schrieffer formalism \cite{bcs} is still
valid in discussions of the doping dependence of the effective
superconducting gap parameter and superconducting transition
temperature, and electron spectral function \cite{ding2}, although
the pairing mechanism is driven by the kinetic energy by
exchanging dressed spin excitations, and other exotic properties,
such as the incommensurate magnetic scattering and commensurate
$[\pi,\pi]$ resonance in the superconducting-state discussed in
the following section, are beyond Bardeen-Cooper-Schrieffer
theory.

\section{Doping and energy dependent incommensurate magnetic
scattering and commensurate resonance}

Experimentally, by virtue of systematic studies using the nuclear
magnetic resonance, and muon spin rotation techniques,
particularly the inelastic neutron scattering, the doping and
energy dependent magnetic excitation spectrum in doped cuprates in
the superconducting-state have been well established: (a) at low
energy, the incommensurate magnetic scattering peaks are shifted
from the antiferromagnetic wave vector $[\pi,\pi]$ to four points
[$(1\pm\delta)\pi, \pi$] and [$\pi,(1\pm\delta)\pi$] with $\delta
$ as the incommensurability parameter \cite{yamada,dai,wakimoto};
(b) then with increasing energy these incommensurate magnetic
scattering peaks are converged on the commensurate [$\pi $,$\pi $]
resonance peak at intermediate energy
\cite{dai,bourges0,bourges,arai}; and (c) well above this
resonance energy, the continuum of magnetic excitations peaked at
the incommensurate positions [$(1\pm\delta)\pi,(1\pm\delta)\pi$]
in the diagonal direction are observed \cite{hayden,tranquada}. It
has been emphasized that the geometry of these incommensurate
magnetic excitations is two-dimensional \cite{hinkov,hayden}.
Although some of these magnetic properties have been observed in
the normal-state, these incommensurate magnetic scattering and
commensurate $[\pi,\pi]$ resonance are the main new feature that
appears into the superconducting-state.

In the charge-spin separation fermion-spin theory, the magnetic
fluctuation is dominated by the scattering of the dressed spins
\cite{feng1,feng7}. Since in the normal-state the dressed spins
move in the dressed holon background, therefore the dressed spin
self-energy (then full dressed spin Green's function) in the
normal-state has been obtained in terms of the collective mode in
the dressed holon particle-hole channel \cite{feng1,feng7}. With
the help of this full dressed spin Green's function in the
normal-state, the incommensurate magnetic scattering and
integrated spin response of doped cuprates in the normal-state
have been discussed \cite{feng1,feng7}, and the results of the
doping dependence of the incommensurability and integrated
dynamical spin susceptibility are consistent with the experimental
results in the normal-state \cite{kastner,yamada,dai}. However, in
the present superconducting-state discussed in section 2, the
antiferromagnetic fluctuation has been incorporated into the
electron off-diagonal Green's function (61) (hence the electron
Cooper pair) in terms of the dressed spin Green's function,
therefore there is a coexistence of the electron Cooper pair and
antiferromagnetic short-range correlation, and then
antiferromagnetic short-range correlation can persist into
superconductivity \cite{feng2}. Moreover, in the
superconducting-state, the dressed spins move in the dressed holon
pair background. In this case, we calculate the dressed spin
self-energy (then the full dressed spin Green's function) in the
superconducting-state in terms of the collective mode in the
dressed holon particle-particle channel, and then give a
quantitative explanation of the incommensurate magnetic scattering
peaks at both low and high energies
\cite{dai,yamada,hayden,tranquada} and commensurate $[\pi,\pi]$
resonance peak at intermediate energy \cite{bourges0,bourges,arai}
in the superconducting-state.

Following our previous discussions for the normal-state case
\cite{feng1,feng7}, the full dressed spin Green's function in the
present superconducting-state is obtained as,
\begin{eqnarray}
D({\bf k},\omega)={1\over D^{(0)-1}({\bf k},\omega)-\Sigma^{(s)}
({\bf k},\omega)},
\end{eqnarray}
with the second order dressed spin self-energy $\Sigma^{(s)}({\bf
k},\omega)$. Within the framework of the equation of motion method
\cite{feng1,feng7}, this dressed spin self-energy in the
superconducting-state with the d-wave symmetry can be obtained
from the dressed holon bubble in the dressed holon
particle-particle channel as,
\begin{eqnarray}
\Sigma_{s}(k)={1\over N^{2}}\sum_{{\bf p,q}} \Lambda({\bf q,p,k})
{1\over \beta}\sum_{iq_{m}}D^{(0)}(q+k) {1\over\beta}\sum_{ip_{m}}
\Im^{\dagger}(p)\Im(p+q),~~~
\end{eqnarray}
where $\Lambda({\bf q,p,k})=[(Zt\gamma_{{\bf k-p}}-Zt'\gamma_{{\bf
k-p}}')^{2}+(Zt\gamma_{{\bf q+p+k}}-Zt'\gamma_{{\bf q+p+k}}')^{2}
]$. This dressed spin self-energy (64) can be evaluated explicitly
in terms of the dressed holon off-diagonal Green's function (32)
and dressed spin mean-field Green's function (20) as,
\begin{eqnarray}
\Sigma_{s}({\bf k},\omega)&=&{1\over N^{2}}\sum_{{\bf p,q}}
\Lambda({\bf q,p,k}){B_{{\bf q+k}}\over \omega_{{\bf
q+k}}}{Z^{2}_{F}\over 4}{\bar{\Delta}^{(a)}_{hZ}({\bf p})
\bar{\Delta}^{(a)}_{hZ}({\bf p+q})\over
E_{{\bf p}}E_{{\bf p+q}}} \nonumber \\
&\times& \left ( {F^{(1)}_{s}({\bf k,p,q})\over \omega^{2}-
(E_{{\bf p}} -E_{{\bf p+q}}+\omega_{{\bf q+k}})^{2}}+
{F^{(2)}_{s}({\bf k,p,q})\over \omega^{2}-(E_{{\bf p+q}} -E_{{\bf
p}}+\omega_{{\bf q+k}})^{2}}
\right . \nonumber \\
&+&{F^{(3)}_{s}({\bf k,p,q})\over \omega^{2}-(E_{{\bf p}}
+ E_{{\bf p+q}} +\omega_{{\bf q+k}})^{2}} \nonumber \\
&+& \left . {F^{(4)}_{s}({\bf k,p,q})\over \omega^{2}- (E_{{\bf
p+q}}+E_{{\bf p}}-\omega_{{\bf q+k}})^{2}} \right ),~~~~~~
\end{eqnarray}
where
\begin{eqnarray}
F^{(1)}_{s}({\bf k,p,q})&=&(E_{{\bf p}}-E_{{\bf p+q}}+
\omega_{{\bf q+k}})\{n_{B}(\omega_{{\bf q+k}})[n_{F}(E_{{\bf p}})
-n_{F}(E_{{\bf p+q}})] \nonumber \\
&-&n_{F}(E_{{\bf p+q}}) n_{F}(-E_{{\bf p}})\},\\
F^{(2)}_{s}({\bf k,p,q})&=&(E_{{\bf p+q}}-E_{{\bf p}}+
\omega_{{\bf q+k}})\{n_{B} (\omega_{{\bf q+k}})[n_{F}(E_{{\bf
p+q}})-n_{F}(E_{{\bf p}})]\nonumber \\
&-&n_{F}(E_{{\bf p}}) n_{F}(-E_{{\bf p+q}})\},\\
F^{(3)}_{s}({\bf k,p,q})&=&(E_{{\bf p}}+E_{{\bf p+q}}
+\omega_{{\bf q+k}})\{n_{B}(\omega_{{\bf q+k}})[n_{F}(-E_{{\bf
p}})- n_{F}(E_{{\bf p+q}})]\nonumber \\
&+&n_{F}(-E_{{\bf p+q}})n_{F}(-E_{{\bf p}})\},\\
F^{(4)}_{s}({\bf k,p,q})&=&(E_{{\bf p}}+E_{{\bf p+q}}-
\omega_{{\bf q+k}})\{n_{B} (\omega_{{\bf q+k}})[n_{F}(-E_{{\bf
p}})-n_{F}(E_{{\bf p+q}})]\nonumber \\
&-&n_{F}(E_{{\bf p+q}})n_{F}(E_{{\bf p}})\}.
\end{eqnarray}
With the help of this full dressed spin Green's function (63), we
can obtain the dynamical spin structure factor in the
superconducting-state with the d-wave symmetry as,
\begin{eqnarray}
S({\bf k},\omega)&=&-2[1+n_{B}(\omega)]{\rm Im}D({\bf k},\omega)
\nonumber \\
&=& -{2[1+n_{B}(\omega)]B^{2}_{{\bf k}}{\rm Im}\Sigma_{s}({\bf k},
\omega)\over [\omega^{2}-\omega^{2}_{{\bf k}}-B_{{\bf k}}{\rm Re}
\Sigma_{s}({\bf k},\omega)]^{2}+[B_{{\bf k}}{\rm Im} \Sigma_{s}
({\bf k},\omega)]^{2}}, ~~~~~
\end{eqnarray}
where ${\rm Im}\Sigma_{s}({\bf k},\omega)$ and ${\rm Re}
\Sigma_{s}({\bf k}, \omega)$ are the imaginary and real parts of
the second order spin self-energy (65), respectively.

\begin{figure}[t]
\begin{center}
\begin{minipage}[h]{100mm}
\epsfig{file=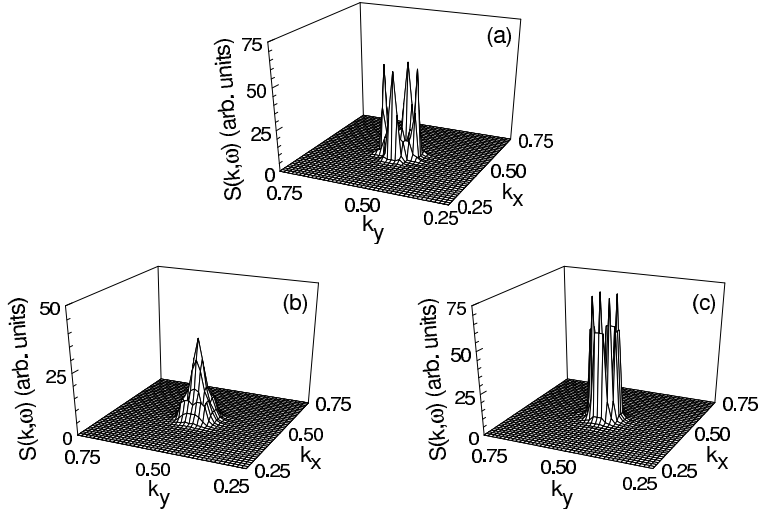, width=100mm}
\end{minipage}
\caption{The dynamical spin structure factor $S({\bf k},\omega)$
in the ($k_{x},k_{y}$) plane at $x_{{\rm opt}}=0.15$ with
$T=0.002J$ for $t/J=2.5$ and $t'/t=0.3$ at (a) $\omega =0.12J$,
(b) $\omega =0.4J$, and (c) $\omega =0.82J$.} \label{F5}
\end{center}
\end{figure}

We are now ready to discuss the doping and energy dependent
magnetic excitation spectrum of doped cuprates in the
superconducting-state. In Fig. 5, we plot the dynamical spin
structure factor $S({\bf k},\omega)$ in the ($k_{x},k_{y}$) plane
at the optimal doping $x_{{\rm opt}}=0.15$ with temperature
$T=0.002J$ for parameters $t/J=2.5$ and $t'/t=0.3$ at energy (a)
$\omega =0.12J$, (b) $\omega =0.4J$, and (c) $\omega =0.82J$. Our
result shows that the distinct feature is the presence of the
incommensurate-commensurate-incommensurate transition in the spin
fluctuation geometry. At low energy, the incommensurate magnetic
scattering peaks are located at $[(1\pm\delta)/2,1/2]$ and
$[1/2,(1\pm\delta )/2]$ (hereafter we use the units of $[2\pi,
2\pi ]$). However, these incommensurate magnetic scattering peaks
are energy dependent, i.e., although these magnetic scattering
peaks retain the incommensurate pattern at $[(1\pm \delta)/2,1/2]$
and $[1/2,(1\pm \delta )/2]$ at low energy, the positions of these
incommensurate magnetic scattering peaks move towards $[1/2,1/2]$
with increasing energy, and then the commensurate $[1/2,1/2]$
resonance peak appears at intermediate energy $\omega_{r}=0.4J$.
This anticipated resonance energy $\omega_{r}=0.4J\approx 40$ meV
\cite{shamoto} is in quantitative agreement with the resonance
energy $\approx 41$ meV observed in the optimally doped
YBa$_2$Cu$_3$O$_{6+y}$ \cite{dai,bourges0,bourges,arai}.
Furthermore, the incommensurate magnetic scattering peaks are
separated again above the commensurate resonance energy, and all
incommensurate magnetic scattering peaks lie on a circle of radius
of $\delta'$ at high energy. In particular, the values of
$\delta'$ at high energy are different from the corresponding
values of $\delta$ at low energy. Although some incommensurate
satellite parallel peaks appear, the main weight of the
incommensurate magnetic scattering peaks is in the diagonal
direction. Moreover, the separation at high energy gradually
increases with increasing energy although the peaks have a weaker
intensity than those below the commensurate resonance energy. To
show this point clearly, we plot the evolution of the magnetic
scattering peaks with energy at $x_{{\rm opt}}=0.15$ in Fig. 6.
For comparison, the experimental result \cite{arai} of
YBa$_2$Cu$_3$O$_{6+y}$ with $y=0.7$ $(x\approx 0.12)$ in the
superconducting-state is also shown in the same figure. The
similar experimental results \cite{bourges,hayden} have also been
obtained for YBa$_2$Cu$_3$O$_{6+y}$ with different doping
concentrations. Our theoretical results show that there is a
narrow energy range for the commensurate resonance peak. The
similar narrow energy range for the commensurate resonance peak
has been observed from experiments \cite{bourges}. Our results
also show that in contrast to the case at low energy, the magnetic
excitations at high energy disperse almost linearly with energy
\cite{arai,hayden,tranquada}. For a better understanding of the
commensurate resonance, we have discussed the doping dependence of
the commensurate resonance, and the result of the resonance energy
$\omega_{r}$ as a function of doping $x-x_{{\rm opt}}$ in
$T=0.002J$ for $t/J=2.5$ and $t'/t=0.3$ is plotted in Fig. 7 in
comparison with the experimental result \cite{bourges0} (inset).
It is shown that in analogy to the doping dependence of the
superconducting transition temperature, the magnetic resonance
energy $\omega_{r}$ increases with increasing doping in the
underdoped regime, and reaches a maximum in the optimal doping,
then decreases in the overdoped regime. These mediating dressed
spin excitations in the superconducting-state are coupled to the
conducting dressed holons (then electrons) under the kinetic
energy driven superconducting mechanism \cite{feng2}, and have
energy greater than the dressed holon pairing energy (then Cooper
pairing energy). Furthermore, we have also made a series of scans
for $S({\bf k},\omega)$ at different temperatures, and found that
those unusual magnetic excitations are present near the
superconducting transition temperature. These results are
quantitatively consistent with the major experimental observations
of doped cuprates in the superconducting-state
\cite{dai,bourges0,bourges,arai,hayden,tranquada}.

\begin{figure}[t]
\begin{center}
\begin{minipage}[h]{85mm}
\epsfig{file=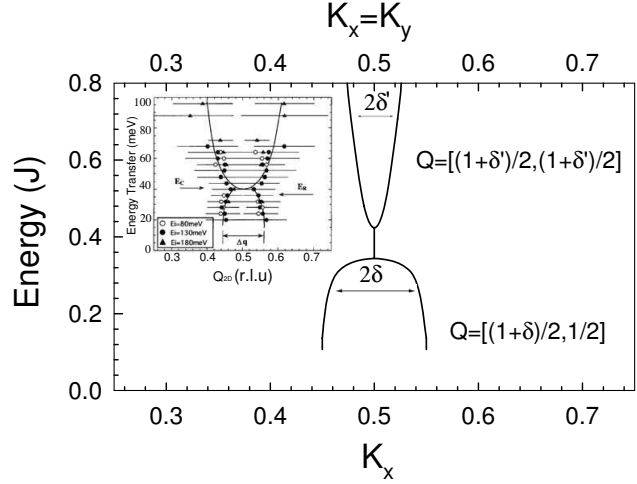, width=85mm}
\end{minipage}
\caption{The energy dependence of the position of the magnetic
scattering peaks at $x_{{\rm opt}}=0.15$ and $T=0.002J$ for
$t/J=2.5$ and $t'/t=0.3$. Inset: the experimental result on
YBa$_{2}$Cu$_{3}$O$_{6.85}$ in the superconducting-state taken
from Ref. \protect\cite{arai}.} \label{F6}
\end{center}
\end{figure}

\begin{figure}[t]
\begin{center}
\begin{minipage}[h]{85mm}
\epsfig{file=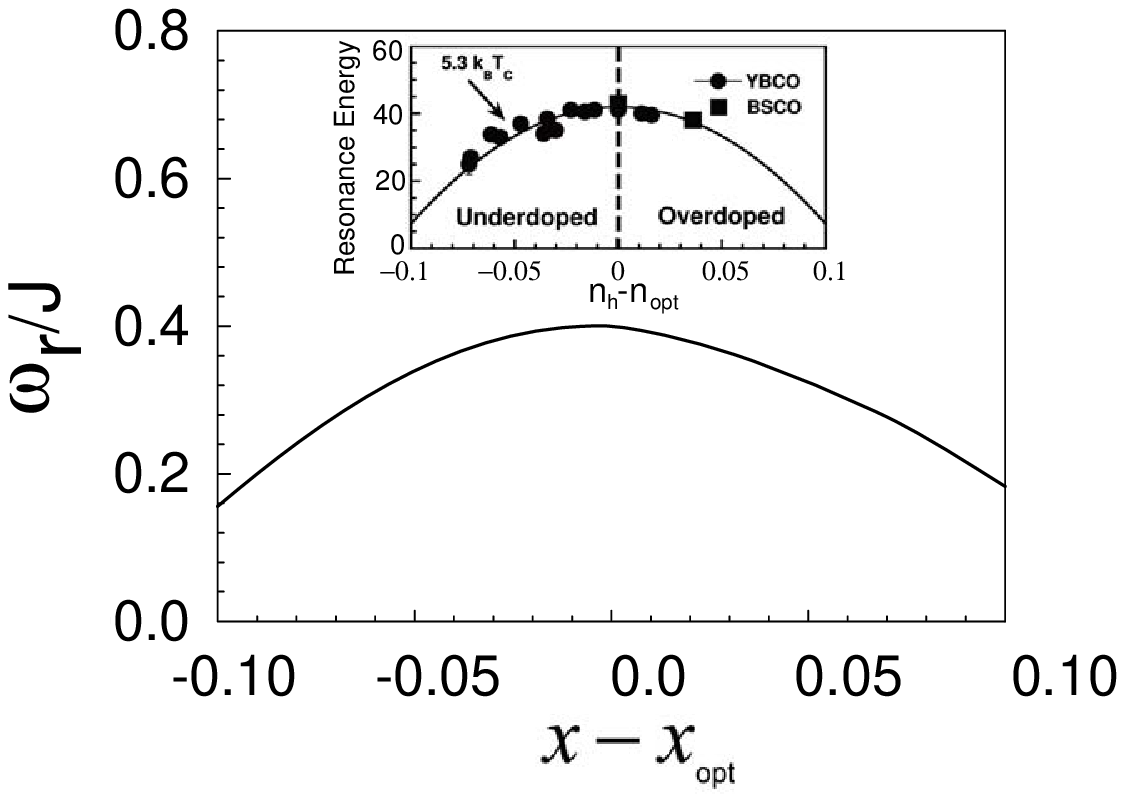, width=85mm}
\end{minipage}
\caption{The resonance energy $\omega_{r}$ as a function of
$x-x_{{\rm opt}}$ with $T=0.002J$ for $t/J=2.5$ and $t'/t=0.3$.
Inset: the experimental result taken from Ref.
\protect\cite{bourges0}.} \label{F7}
\end{center}
\end{figure}

The physical interpretation to the above obtained results can be
found from the property of the renormalized dressed spin
excitation spectrum $\Omega^{2}_{{\bf k}}=\omega^{2}_{{\bf k}}+
{\rm Re}\Sigma^{(s)}({\bf k},\Omega_{{\bf k}})$ in Eq. (70). Since
both mean-field dressed spin excitation spectrum $\omega_{{\bf
k}}$ and dressed spin self-energy function $\Sigma^{(s)}({\bf
k},\omega)$ in Eq. (65) are strong doping and energy dependent,
this leads to that the renormalized dressed spin excitation
spectrum also is strong doping and energy dependent. The dynamical
spin structure factor in Eq. (70) has a well-defined resonance
character, where $S({\bf k},\omega)$ exhibits peaks when the
incoming neutron energy $\omega$ is equal to the renormalized spin
excitation, i.e.,
\begin{eqnarray}
W({\bf k}_{c},\omega)\equiv [\omega^{2}- \omega_{{\bf
k}_{c}}^{2}-B_{{\bf k}_{c}}{\rm Re}\Sigma^{(s)}({\bf k}_{c},
\omega)]^{2}= [\omega^{2}-\Omega_{{\bf k}_{c}}^{2}]^{2} \sim 0,
\end{eqnarray}
for certain critical wave vectors ${\bf k}_{c}={\bf k}^{(L)}_{c}$
at low energy, ${\bf k}_{c}={\bf k}^{(I)}_{c}$ at intermediate
energy, and ${\bf k}_{c}={\bf k}^{(H)}_{c}$ at high energy, then
the weight of these peaks is dominated by the inverse of the
imaginary part of the dressed spin self-energy $1/{\rm
Im}\Sigma^{(s)}({\bf k}^{(L)}_{c},\omega)$ at low energy, $1/{\rm
Im}\Sigma^{(s)}({\bf k}^{(I)}_{c},\omega)$ at intermediate energy,
and $1/{\rm Im}\Sigma^{(s)}({\bf k}^{(H)}_{c},\omega)$ at high
energy, respectively. In the normal-state \cite{feng1,feng7}, the
dressed holon energy spectrum has one branch $\xi_{{\bf k}}$,
while in the present superconducting-state, the dressed holon
quasiparticle spectrum has two branches $\pm E_{{\bf k}}$, this
leads to that the dressed spin self-energy function
$\Sigma^{(s)}({\bf k},\omega)$ in Eq. (65) is rather complicated,
where there are four terms in the right side of Eq. (65). In
comparison with the normal-state case \cite{feng1,feng7}, the
contribution for the first and second terms in the right side of
the dressed spin self-energy (65) comes from the lower band
$-E_{{\bf k}}$ of the dressed holon quasiparticle spectrum like
the normal-state case, while the contribution for the third and
fourth terms in the right side of the dressed spin self-energy
(65) comes from the upper band $E_{{\bf k}}$ of the dressed holon
quasiparticle spectrum. During the above calculation, we find that
the mode which opens downward and gives the incommensurate
magnetic scattering at low energy is mainly determined by the
first and second terms in the right side of the dressed spin
self-energy (65), while the mode which opens upward and gives the
incommensurate magnetic scattering at high energy is essentially
dominated by the third and fourth terms in the right side of the
dressed spin self-energy (65), then two modes meet at the
commensurate $[1/2,1/2]$ resonance at intermediate energy. This
means that within the framework of the kinetic energy driven
superconductivity, as a result of self-consistent motion of the
dressed holon pairs and spins, the incommensurate magnetic
scattering at both low and high energies and commensurate
resonance at intermediate energy are developed. This reflects that
the low and high energy spin excitations drift away from the
antiferromagnetic wave vector, or the zero point of $W({\bf
k}_{c}, \omega)$ is shifted from $[1/2,1/2]$ to ${\bf k}_{c}={\bf
k}^{(L)}_{c}$ at low energy and ${\bf k}_{c}={\bf k}^{(H)}_{c}$ at
high energy. With increasing energy from low energy or decreasing
energy from high energy, the spin excitations move towards to
$[1/2,1/2]$, i.e., the zero point of $W({\bf k}_{c},\omega)$ in
${\bf k}_{c}={\bf k}^{(L)}_{c}$ at low energy or ${\bf k}_{c}={\bf
k}^{(H)}_{c}$ at high energy turns back to $[1/2,1/2]$, then the
commensurate $[1/2,1/2]$ resonance appears at intermediate energy.
To show this point clearly, the function $W({\bf k}, \omega)$ in
$x_{{\rm opt}} =0.15$ for $t/J=2.5$ and $t'/t=0.3$ with $T=0.002J$
from (a) ${\bf k}_{1} =[(1- \delta)/2, 1/2]$ via ${\bf k}_{2}
=[1/2,1/2]$ to ${\bf k}_{3}= [(1+\delta)/2,1/2]$ at $\omega=0.12J$
(solid line) and $\omega= 0.4J$ (dashed line), and (b) ${\bf
k}_{4}= [(1- \delta')/2,(1- \delta')/2]$ via ${\bf k}_{2}=
[1/2,1/2]$ to ${\bf k}_{5}=[(1+ \delta')/2,(1+\delta')/2]$ at
$\omega=0.4J$ (solid line) and $\omega=0.82J$ (dashed line) is
plotted in Fig. 8, where there is a strong angular dependence with
actual minima in $[(1-\delta)/2, 1/2]$ and $[1/2,(1-\delta)/2]$,
$[1/2,1/2]$, and $[(1-\delta')/2,( 1-\delta')/2]$ and
$[(1+\delta')/2, (1+ \delta')/2]$ for low, intermediate, and high
energies, respectively. These are exactly positions of the
incommensurate magnetic scattering peaks at both low and high
energies and commensurate resonance peak at intermediate energy
determined by the dispersion of very well defined renormalized
spin excitations. Since the essential physics is dominated by the
dressed spin self-energy renormalization due to the dressed holon
bubble in the dressed holon particle-particle channel, then in
this sense the mobile dressed holon pairs (then the electron
Cooper pairs) are the key factor leading to the incommensurate
magnetic scattering peaks at both low and high energies and
commensurate resonance peak at intermediate energy, i.e., the
mechanism of the incommensurate magnetic scattering and
commensurate resonance in the superconducting-state is most likely
related to the motion of the dressed holon pairs (then the
electron Cooper pairs). This is why the positions of the
incommensurate magnetic scattering peaks and commensurate
resonance peak in the superconducting-state can be determined in
the present study within the $t$-$t'$-$J$ model based on the
kinetic energy driven superconducting mechanism, while the dressed
spin energy dependence is ascribed purely to the self-energy
effects which arise from the the dressed holon bubble in the
dressed holon particle-particle channel.

\begin{figure}[t]
\begin{center}
\begin{minipage}[h]{130mm}
\epsfig{file=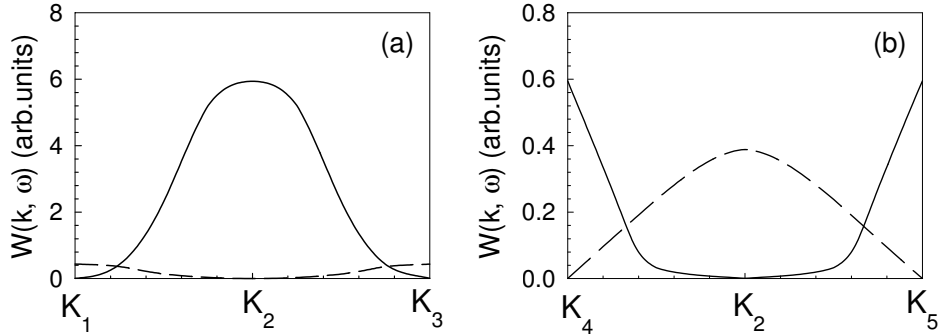, width=125mm}
\end{minipage}
\caption{Function $W({\bf k},\omega)$ in $x_{{\rm opt}}=0.15$ for
$t/J=2.5$ and $t'/t=0.3$ with $T=0.002J$ from (a) ${\bf k}_{1}=
[(1-\delta)/2,1/2] $ via ${\bf k}_{2}=[1/2,1/2]$ to ${\bf k}_{3}
=[(1+\delta)/2,1/2]$ at $\omega=0.12J$ (solid line) and
$\omega=0.4J$ (dashed line), and (b) ${\bf k}_{4}=
[(1-\delta')/2,(1-\delta')/2]$ via ${\bf k}_{2}= [1/2,1/2]$ to
${\bf k}_{5}=[(1+\delta')/2,(1+\delta')/2]$ at $\omega=0.4J$
(solid line) and $\omega=0.82J$ (dashed line).} \label{F8}
\end{center}
\end{figure}

\section{The charge asymmetry in superconductivity of hole and
electron doping}

Superconductivity emerges when charge carriers, holes or
electrons, are doped into the parent compound of cuprates
\cite{bednorz,tokura}. Both hole-doped and electron-doped cuprate
superconductors have the layered structure of the square lattice
of the CuO$_{2}$ plane separated by insulating layers
\cite{kastner,tokura}. It has been found from experiments that
only an approximate symmetry in the phase diagram exists about the
zero doping line between the electron and hole doping \cite{sawa}.
For the hole-doped case \cite{bednorz,kastner}, the
antiferromagnetic long-range order disappears rapidly with doping,
and is replaced by a disordered spin liquid phase, then the
systems become superconducting over a wide range of the hole
doping concentration, around the optimal $x\sim 0.15$
\cite{tallon}. However, the antiferromagnetic long-range order
survives until superconductivity appears over a narrow range of
the electron doping concentration around the optimal $x\sim 0.15$
in the electron-doped case, where the maximum achievable
superconducting transition temperature is much lower than that in
the hole-doped case \cite{tokura,tokura1,peng}. Although this
electron-hole asymmetry is observed in the phase diagram
\cite{kastner,sawa}, the charge carrier Cooper pairs in both
optimally electron- and hole-doped cuprate superconductors have a
dominated d-wave symmetry \cite{tsuei,martindale,armitage,tsuei2}.
Since the strong electron correlation is common for both
hole-doped and electron-doped cuprate superconductors, many of the
physical properties of electron-doped cuprate superconductors
resemble that of the hole-doped case. These show that both hole-
and electron-doped cuprate superconductors have similar underlying
superconducting mechanism. In this section, we study the charge
asymmetry in superconductivity of hole- and electron-doped cuprate
superconductors. We show that superconductivity appears over a
narrow range of the electron doping concentration in the
electron-doped case, and the maximum achievable superconducting
transition temperature in the optimal doping is lower than that of
the hole-doped case due to the electron-hole asymmetry.

\begin{figure}[t]
\begin{center}
\begin{minipage}[h]{130mm}
\epsfig{file=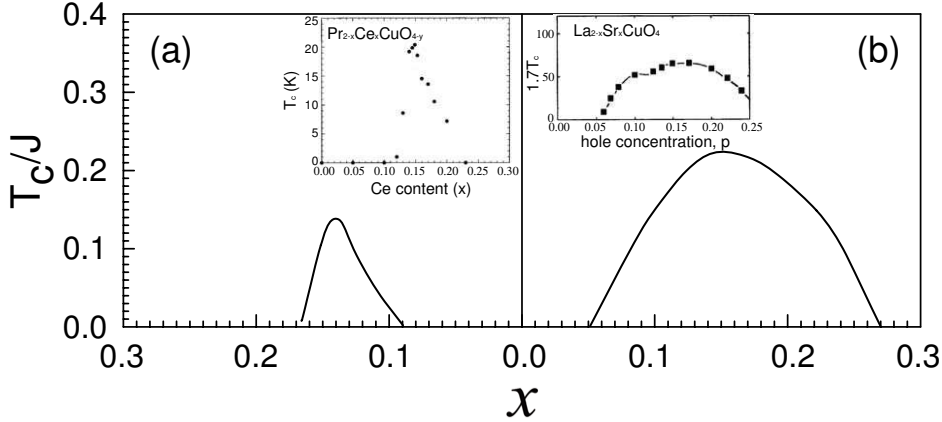, width=125mm}
\end{minipage}
\caption{The superconducting transition temperature as a function
of the doping concentration with (a) $t/J=-2.5$ and $t'/t=0.3$ for
the electron doping and (b) $t/J=2.5$ and $t'/t=0.3$ for the hole
doping. Inset: the corresponding experimental results of
Pr$_{2-x}$Ce$_{x}$CuO$_{4-y}$ taken from Ref. \protect\cite{peng}
and La$_{2-x}$Sr$_{x}$CuO$_{4}$ from Ref. \protect\cite{tallon}}
\label{F9}
\end{center}
\end{figure}

For discussing the case of the electron doping, we can perform a
particle-hole transformation $C_{i\sigma}\rightarrow
C^{\dagger}_{i-\sigma}$ for the $t$-$t'$-$J$ model (1), so that
the difference between hole and electron doping is expressed as
the sign difference of the hopping parameters \cite{feng4}, i.e.,
$t>0$ and $t'>0$ for the hole doping and $t<0$ and $t'<0$ for the
electron doping, then the $t$-$t'$-$J$ model (1) in both hole- and
electron-doped cases is always subject to an important on-site
local constraint to avoid the double occupancy. Moreover, it has
been shown that the absolute values of $t$ and $t'$ are almost
same for both hole- and electron-doped cuprates \cite{hybertson}.
In this case, we have followed the discussions in section 2 for
the hole-doped case, and performed a calculation for the case of
the electron doping. In Fig. 9(a), we plot the superconducting
transition temperature in the case of the d-wave symmetry as a
function of the electron doping concentration for $t/J=-2.5$ and
$t'/t=0.3$ in comparison with the corresponding experimental
result of Pr$_{2-x}$Ce$_{x}$CuO$_{4-y}$ \cite{peng} (inset). For
comparison, the corresponding result of the hole doping in the
case of the d-wave symmetry in Fig. 2 is also replotted in Fig.
9(b). Our results indicate that in analogy to the phase diagram of
the hole-doped case, superconductivity appears over a narrow range
of doping in the electron-doped side, where the superconducting
transition temperature increases sharply with increasing doping in
the underdoped regime, and reaches a maximum in the optimal doping
$x_{{\rm opt}}\approx 0.14$, then decreases sharply with
increasing doping in the overdoped regime. However, the maximum
achievable superconducting transition temperature in the optimal
doping in the electron-doped case is much lower than that of the
hole-doped case due to the electron-hole asymmetry. Using an
reasonably estimative value of $J\sim 800$K to 1200K for the
electron-doped cuprate superconductors, the superconducting
transition temperature in the optimal doping is T$_{c}\approx
0.136J \approx 108{\rm K}\sim 163{\rm K}$, in semiquantitative
agreement with the corresponding experimental data
\cite{tokura,peng}. The essential physics of the doping dependent
superconducting transition temperature in the electron-doped case
is almost the same as in the hole-doped side, and detailed
explanations have been given in section 2. On the other hand, it
has been shown \cite{hybertson} that the antiferromagnetic
long-range order can be stabilized by the $t'$ term for the
electron-doped case, which may lead to the charge carrier's
localization over a broader range of doping, this is also why
superconductivity appears over a narrow range of doping in
electron-doped cuprate superconductors. Furthermore, we have
discussed the spin response of the electron-doped cuprates in the
superconducting-state, and the new feature of the spin response in
the electron-doped cuprates in the superconducting-state is the
energy dependence of the commensurate resonance peak at both low
and intermediate energies and incommensurate magnetic scattering
peaks at high energy. These and related theoretical results will
be presented elsewhere.

\section{Conclusion}

Within the $t$-$t'$-$J$ model, we have discussed the physical
properties of doped cuprates in the superconducting-state based on
the charge-spin separation fermion-spin theory. It is shown that
the superconducting-state is controlled by both superconducting
gap parameter and single particle coherence, and is the
conventional Bardeen-Cooper-Schrieffer like, so that some of the
basic Bardeen-Cooper-Schrieffer formalism \cite{bcs} is still
valid in quantitatively reproducing of the doping dependence of
the superconducting gap parameter \cite{wen}, the doping
dependence of the superconducting transition temperature
\cite{tallon}, and the electron spectral function in the
superconducting-state \cite{ding,well}, although the
superconducting Cooper pairing mechanism is driven by the kinetic
energy by exchanging dressed spin excitations \cite{feng2}, and
other exotic properties are beyond Bardeen-Cooper-Schrieffer
theory. Although the symmetry of the superconducting-state is
doping dependent, the superconducting-state has the d-wave
symmetry in a wide range of doping. Within this d-wave
superconducting-state, we have performed a systematic calculation
for the dynamical spin structure factor of doped cuprates in the
superconducting-state in terms of the collective mode in the
dressed holon particle-particle channel, and quantitatively
reproduced all main features found in the inelastic neutron
scattering experiments on cuprate superconductors, including the
energy dependence of the incommensurate magnetic scattering at
both low and high energies \cite{hayden,arai,tranquada} and
commensurate $[\pi,\pi ]$ resonance at intermediate energy
\cite{bourges0,bourges}. In particular, we have shown that the
unusual incommensurate magnetic excitations at high energy have
energies greater than the dressed holon pairing energy (then
superconducting Cooper pairing energy), and are present at the
superconducting transition temperature. Furthermore, we have
studied the charge asymmetry of superconductivity in the hole and
electron doping, and show that in analogy to the phase diagram of
the hole-doped case, superconductivity appears over a narrow range
of the electron doping concentration in the electron-doped side,
and the maximum achievable superconducting transition temperature
in the optimal doping in the electron-doped case is much lower
than that of the hole-doped side due to the electron-hole
asymmetry. Our present study also shows that the effect of the
additional second neighbor hopping $t'$ is to enhance the d-wave
superconducting pairing correlation, and suppress the s-wave
superconducting pairing correlation.

Within this framework of the kinetic energy driven
superconductivity, we \cite{guo20} have studed the electronic
structure of cuprate superconductors. It is shown that the
spectral weight of the electron spectrum in the antinodal point of
the Brillouin zone decreases as the temperature is increased. With
increasing the doping concentration, this spectral weigh
increases, while the position of the sharp superconducting
quasiparticle peak moves to the Fermi energy. In analogy to the
normal-state case, the superconducting quasiparticles around the
antinodal point disperse very weakly with momentum. Our results
also show that the striking behavior of the superconducting
coherence of the quasiparticle peaks is intriguingly related to
the strong coupling between the superconducting quasiparticles and
collective magnetic excitations.

\begin{center}
{\bf Acknowledgements}
\end{center}

The author would like to thank Dr. Huaiming Guo, Dr. Yu Lan, Dr.
Yiny Liang, Dr. Bin Liu, and Professor Y.J. Wang for the helpful
discussions. This work was supported by the National Natural
Science Foundation of China under Grant Nos. 10125415 and
90403005, and the Grant from Beijing Normal University.



\end{document}